\documentclass{article}

\usepackage{arxiv}

\usepackage[utf8]{inputenc} 
\usepackage[T1]{fontenc}    
\usepackage{hyperref}       
\usepackage{url}            
\usepackage{booktabs}       
\usepackage{amsfonts}       
\usepackage{nicefrac}       
\usepackage{microtype}      
\usepackage{lipsum}
\usepackage{graphicx}
\graphicspath{ {./images/} }

\usepackage{cite}
    
\usepackage{bbding}
\usepackage{pifont}
\usepackage[nolist]{acronym}
\usepackage{booktabs}
\usepackage{amssymb}
\usepackage{amsmath}
\usepackage{wasysym}
\usepackage[dvipsnames]{xcolor}
\usepackage{amsmath,amssymb,amsfonts}
\usepackage{booktabs}
\usepackage[ruled,vlined,linesnumbered]{algorithm2e}
\usepackage{array}
\usepackage{tabularx}
\usepackage{comment}
\usepackage{colortbl}
\usepackage{float}
\usepackage{xfrac}

\usepackage[inline]{enumitem}
\usepackage{caption}
\usepackage{subcaption}
\usepackage{multirow}
\usepackage{adjustbox}
\usepackage{algorithmicx}
\usepackage{textcomp}
\usepackage{xcolor}
\usepackage{hyperref}
\usepackage{color,soul}
\usepackage{caption}
\newcommand{\Revised}{\color{black}}

\title{iOn-Profiler: intelligent Online multi-objective VNF Profiling with Reinforcement Learning}

\author{
  Xenofon Vasilakos \\
  \texttt{xenofon.vasilakos@bristol.ac.uk} \\
  Smart Internet Lab, \\
  Department of Electrical \& Electronic Engineering,\\ 
  University of Bristol, BS8 1UB, UK
  \And
  Shadi Moazzeni \\
  \texttt{shadi.moazzeni@bristol.ac.uk} \\
  Smart Internet Lab, \\
  Department of Electrical \& Electronic Engineering,\\ 
  University of Bristol, BS8 1UB, UK
  \And
  Anderson Bravalheri \\
  \texttt{a.bravalheri@bristol.ac.uk} \\
  Smart Internet Lab, \\
  Department of Electrical \& Electronic Engineering,\\ 
  University of Bristol, BS8 1UB, UK
  \And
  Pratchaya Jaisudthi \\
  \texttt{yo19607@bristol.ac.uk} \\
  Smart Internet Lab, \\
  Department of Electrical \& Electronic Engineering,\\ 
  University of Bristol, BS8 1UB, UK
  \And
  Reza Nejabati \\
  \texttt{reza.nejabati@bristol.ac.uk} \\
  Smart Internet Lab, \\
  Department of Electrical \& Electronic Engineering,\\ 
  University of Bristol, BS8 1UB, UK
  \And
  Dimitra Simeonidou \\
  \texttt{dimitra.simeonidou@bristol.ac.uk} \\ 
  Smart Internet Lab, \\
  Department of Electrical \& Electronic Engineering,\\ 
  University of Bristol, BS8 1UB, UK
}

\newcommand{\rl}{\ac{rl} }

\newcommand{\vnf}{\ac{vnf} }

\newcommand{\iperf}{iPerf }
\newcommand{\snort}{Snort }
\newcommand{\vfw}{\ac{vfw} }

\newcommand{\lc}{\ac{lc} }

\newcounter{Qcountnum}

\newcommand\nq[3]{
	{
      \stepcounter{Qcount\roman{Qcountnum}}
      \textbf{\color{red}\#Quest.\_\csname theQcount\roman{Qcountnum}\endcsname } \color{red}$<$From:#1 To:#2$>$: {\color{red}\hl{#3}}
    }
}

\newcounter{Rcountnum}

\newcommand\nr[1]{
	{\color{red}
      \stepcounter{Rcount\roman{Rcountnum}}
      \#note\_\csname theRcount\roman{Rcountnum}\endcsname:[[#1]]}}

\newcounter{Bcountnum}

\newcommand\nb[1]{
	{\color{brown}
      \stepcounter{Bcount\roman{Bcountnum}}
      \#note\_\csname theBcount\roman{Bcountnum}\endcsname:[[#1]]}}

\newcounter{Mcountnum}

\newcommand\nshadim[1]{
	{\color{magenta}
      \stepcounter{Mcount\roman{Mcountnum}}
      \#note\_\csname theMcount\roman{Mcountnum}\endcsname:[[#1]]}}

\DeclareMathOperator*{\argmax}{argmax}

\DeclareMathOperator*{\rand}{rand}
    
\begin{document}


\begin{acronym} 


\acro{ids}[IDS]{Intrusion Detection System}
\acro{vm}[VM]{Virtual Machine}
\acro{bss}[BSS]{Binary Section Search}
\acro{vfw}[vFW]{virtual FireWall}


\acro{nf}[NF]{Network Function}

\acro{vnf}[VNF]{Virtualised Network Function}
\acrodefplural{vnf}[VNFs]{Virtualised Network Functions}
\acro{nfv}[NFV]{Network Function Virtualisation}
\acro{nat}[NAT]{Network Address Translator}
\acro{dns}[DNS]{Domain Name System}
\acro{ns}[NS]{Network Service}
\acro{sfc}[SFC]{Service Function Chain}
\acro{rtt}[RTT]{Round Trip Time}

\acro{osm}[OSM]{Open Source MANO}
\acro{mano}[MANO]{MANagement and Orchestration}

\acro{vna}[VNA]{Virtual Network Applications}
\acro{sdn}[SDN]{Software-Defined Networking}
\acro{mln}[MLN]{Machine Learning in Networking}
\acro{zsm}[ZSM]{Zero-touch network and Service Management}
\acro{mec}[MEC]{Multi-access Edge Computing}

\acro{soo}[SOO]{Single-Objective Optimisation}
\acro{moo}[MOO]{Multi-Objective Optimisation}
\acro{etsi}[ETSI]{European Telecommunications Standards Institute}
\acro{nsd}[NSD] {Network Service Descriptor}
\acro{nfvi}[NFVI]{NFV Infrastructure}

\acro{ml}[ML]{Machine Learning}
\acro{rl}[RL]{Reinforcement Learning}
\acro{mdp}[MDP]{Markov Decision Process}
\acro{mab}[MAB]{Multi-Armed Bandit}
\acro{rf}[RF]{Random Forest}
\acro{ann}[ANN]{Artificial Neural Network}

\acro{sl}[SL]{Supervised Learning}
\acro{ul}[UL]{Unsupervised Learning}
\acro{tsf}[TSF]{Time Series Forecasting}
\acro{dt}[DT]{Decision Tree}
\acro{lr}[LR]{Linear Regression}
\acro{svr}[SVR]{Support Vector Regression}
\acro{krr}[KRR]{Kernel Ridge Regression}
\acro{lar}[LAR]{Lasso Regression}
\acro{knnr}[KNNR]{K-Nearest Neighbors Regression}
\acro{lstm}[LSTM]{Long-Short Term Memory}

\acro{nap}[NAP]{Novel Autonomous Profiling}

\acro{ar}[AR]{Autoregression}
\acro{ma}[MA]{Moving Average}
\acro{es}[ES]{Exponential Smoothing}
\acro{arma}[ARMA]{Autoregressive Moving Average}

\acro{vim}[VIM]{Virtualised Infrastructure Manager}
\acro{mano}[MANO]{MANagement and Orchestration}
\acro{osm}[OSM]{Open Source MANO}
\acro{osm-mano}[OSM MANO]{Open Source MANagement and Orchestration framework}
\acro{nfvo}[NFVO]{NFV Orchestrator}


\acro{lc}[LC]{Link Capacity}
\acro{or}[OR]{Output Rate}
\acro{ir}[IR]{Input Rate}

\acro{kpi}[KPI]{Key Performance Indicator}
\acro{poc}[PoC]{Proof-of-Concept}
\acro{sla}[SLA]{Service Level Agreement}
\acro{slas}[SLAs]{Supervised Learning Algorithms}
\acro{ue}[UE]{User Equipment}

\acro{rmse}[RMSE]{Root Mean Square Error}
\acro{mape}[MAPE]{Mean Absolute Percentage Error}
\acro{qos}[QoS]{Quality-of-Service}
\acro{qoe}[QoE]{Quality-of-Experience}
\acro{ete}[e2e]{end-to-end}

\acro{cpu}[CPU]{Central Processor Unit}
\acro{vcpu}[vCPU]{virtual Central Processor Unit}
\acro{gpu}[GPU]{Graphical Processing Unit}
\acro{tpu}[TPU]{Tensor Processing Unit}
\acro{mlp}[MLP]{Multi-Layer Perceptron}

\end{acronym}

\maketitle
\begin{abstract}
Leveraging the potential of Virtualised Network Functions (VNFs) requires a clear understanding of the link between \textit{resource consumption} and \textit{performance}. The current state of the art tries to do that by utilising Machine Learning (ML) and specifically Supervised Learning (SL) models for given network environments and VNF types assuming single-objective optimisation targets. Taking a different approach, \textit{iOn-Profiler} poses a \textit{novel VNF profiler} optimising \textit{multi-}resource type allocation and performance objectives using adapted Reinforcement Learning (RL).
Our approach can meet Key Performance Indicator (KPI) targets while minimising multi-resource type consumption and optimising the VNF output rate compared to existing single-objective solutions. Our experimental evaluation with \textit{three} \textit{real-world} VNF types over a total of \textit{39 study scenarios} (13 per VNF),
for three resource types (virtual CPU, memory, and network link capacity),
verifies the accuracy of resource allocation predictions and corresponding successful profiling decisions via a benchmark comparison between our RL model and SL models. We also conduct a complementary exhaustive search-space study revealing that different resources impact performance in varying ways per VNF type, implying the \emph{necessity} of multi-objective optimisation, individualised examination per VNF type, and adaptable online profile learning, such as with the \textit{autonomous online learning} approach of iOn-Profiler.
\end{abstract}

\section{Introduction}

The rise of Cloud computing, \ac{sdn} and \ac{nfv} have caused a paradigm shift from traditional networking based on specialised hardware to utilising general-purpose programmable hardware as a \textit{resource for running \acp{vnf}}. This change has simplified the design, deployment, and management of network services, with network-based service providers offering \acp{sla} to their customers that outline performance requirements \textit{and} \ac{kpi} levels (e.g., throughput, packet loss, response time, processing latency, and so forth). 

{\Revised 
At the core of achieving \ac{sla} goals lies the essential \textit{process of} \textit{\ac{vnf} profiling}. 
It involves the systematic analysis and characterisation of different \acp{vnf} within a programmable \ac{sdn} environment. 
The primary objective is to understand each \ac{vnf}'s individual resource requirements, performance expectations and operational behaviour by 
discovering the \textit{relationship} between resource configuration and performance. 
This knowledge enables service providers to decide the \textit{(i) optimal allocation} of network and computation resources such as CPU or bandwidth, for each \ac{vnf} 
instance, while \textit{(ii)} ensuring \textit{adherence} to predefined \ac{kpi} thresholds after \ac{sla} goals. 
\ac{vnf} profiling is undertaken by a ``VNF profiler'' and the resulting profile describes a discovered reciprocal mapping between optimised resource 
allocations and the \ac{kpi} thresholds for the respective \ac{vnf}, which enables knowing the expected performance after allocating resources and vice versa.
}
 
In the context of contemporary networks such as 5G and future 6G, attention to profiling is driven by its significance for \ac{nfv} \ac{mano} systems. 
The latter can use \ac{vnf} profiles to instantiate \acp{ns} by adapting optimised resource configurations. Moreover, profiles can be used to optimise the life-cycle management of running services. As an example, 
the 5G-VIOS~\cite{vios_pro} \textit{common interfacility orchestration platform} leverages autonomously generated~\cite{NAP} profiling models to deploy and orchestrate inter-edge \acp{ns} across multiple domains and 
facilities by (i) \textit{autonomously} assigning optimised resource configurations to inter-edge \acp{ns} while also (ii) exposing corresponding performance profiles.

{\Revised
The current work presents \textit{iOn-Profiler}, an online \ac{vnf} profiler that leverages adaptive \ac{rl}. In summary, our most significant and novel contributions are:}
\begin{enumerate}
    \item \textbf{{\Revised Online,} 
    multi-objective optimisation profiling:}
    {\Revised
    We investigate \ac{rl}-based adaptive \ac{vnf} profiling for minimising the use of
    both compute and network resources,
    as well as finding the Optimum~\ac{or}. The latter stands for the output rate achieved by the profiled \ac{vnf} under an optimal (i.e., minimum) resource configuration that meets \ac{kpi} targets.}
    
    \item \textbf{{\Revised Pragmatic} \ac{vnf} case studies:} We consider \textit{three} \textit{pragmatic} \acp{vnf} in our experimental study, namely a \ac{vfw} and two different modes of the Snort~\cite{snort_modes} open source intrusion prevention system (Inline and Passive modes). {\Revised Besides pragmatic, these \acp{vnf} span both dissimilar and similar features, allowing to assess functionality footprint on resulting profiles.} 
    
    \item \textbf{Oracle exhaustive search:} We conduct an exhaustive search of resource-to-\ac{kpi} combinations for all \acp{vnf} involving all resource types to establish an all-possible performance knowledge and understanding of the impact importance of different resource types on the performance of different \ac{vnf} types.
    %
    Then we utilise this Oracle-gained insight to carefully explore and tune the \ac{rl} reward function parameters of iOn-Profiler.

    \item \textbf{Extensive experimental analysis:} Overall, our evaluation study highlights that different resources impact \ac{vnf} performance in distinct ways. Besides Oracle search, this conclusion is also established through the analysis of each \ac{vnf} type's Pareto front over a total of \textit{39 scenarios} (13 per each of 3 \ac{vnf} types). 
    Our highlight results and conclusions include: 
    \begin{itemize}
        \item Multi-objective optimisation is \textit{necessary} for proper \ac{vnf} profiling.
        
        \item There is a strong requirement for studying each \ac{vnf} \textit{type} and \textit{mode of operation} \textit{individually} such as demonstrated for Snort (Passive \textit{vs.} Inline modes).
       
        \item Online learning is \textit{significant}, as fixed \ac{sl} models lack adaptability to dynamics.
    \end{itemize}
\end{enumerate}

{\Revised The rest of this article is organised as follows. 
Section~\ref{sec:background-r01} provides the reader with the necessary background context.} 
Section~\ref{sec:profiler-design} discusses the design of iOn-Profiler. 
The experimental setup, resource and model configurations are described in Sec.~\ref{sec:eval-setup}.
Our experimental evaluation is presented in Sec.~\ref{sec:eval} followed by our future work and conclusion in Sec.~\ref{sec:conclusion}.

{\Revised  
\section{Background \& Motivation in Intelligent VNF profiling}\label{sec:background-r01}

We discuss the essential background context, encompassing the problem statement, the specific objectives set for the proposed iOn-Profiler solution, and an in-depth analysis of the state of the art in intelligent \ac{vnf} profiling. 

   
\subsection{Problem statement \& utilising machine learning}\label{sec:probState}

{\Revised

Let $I$ be the set of all considered \textit{resource types} $i \in I$, and $K$ be the set of every considered \ac{kpi} type $k \in K$. 
Also,  Let $\mathbf{x} = \{x_1, x_2, ... x_v\}$ be the decision vector of \ac{kpi} threshold targets for allocating resources. Each threshold target $x_k$ corresponds to \ac{kpi} $k$ and can get a value only from the partition set $T_k$ defined below. Last, $\mathbf{x} \in \mathbf{X}$, where set $\mathbf{X}$ is the feasible set of decision vectors.
Let the set $T_k = \{\tau_1, \tau_2, ... \tau_{\omega}\}$ be an partition set of considered performance thresholds for \ac{kpi} type $k$.
Last, let $f_i(\mathbf{x})$ be the allocated amount for resource type $i$, given the \ac{kpi} threshold targets $\mathbf{x}$. 
We define the following \textit{multi-objective} optimisation problem:  
{
    \begin{equation}
        \begin{aligned}
            &\underset{\mathbf{x} \in \mathbf{X}}{\mathrm{min}} \, (f_1(\mathbf{x}), f_2(\mathbf{x}), \ldots, f_n(\mathbf{x})) \\
            &\text{Subject to:} \quad m_k \geq \tau_k,\,  \forall k
        \end{aligned}
    \end{equation}
}
%

\noindent To consider maximum \ac{kpi} thresholds as well (e.g., for packet drop rate), we adopt appropriate minimum and maximum threshold constraints $\tau^{min}_k$ and $\tau^{max}_k$ and redefine the problem constraints as: $m_k \geq \tau^{min}_k,\,  \forall k$ and $m_k \leq \tau_k^{\max},\,  \forall k$.
}

As detailed in Sec.~\ref{sec:related-works}, \ac{ml} poses a dominant trend in the \ac{vnf} profiling literature due to its adaptability to complex environments. 
Compared to other types of prominent works based on linear programming and heuristics (e.g., ~\cite{luizelli2015piecing, fang2016}) \ac{ml} solutions delve deeper into \ac{vnf}-to-resource specifics, with core challenges captured \textit{better}:
    First, \ac{ml} can capture better network and service \textit{dynamics}, particularly regarding 5G and 6G programmable networks due to their agility. 
    Second, they can do so \textit{within practical time-scales} despite the \textit{NP-hardness}~\cite{ztorch} of the underlying optimisation problem, 
by \textit{converging} towards optimised configurations involving different resources and subject to \ac{kpi} targets.

Prominent examples of \ac{ml} models used for profiling include Linear Regression~\cite{montgomery2021introduction}, \ac{knnr}\cite{gou2019generalized}, Interpolation\cite{verma2019interpolation}, \acp{ann}\cite{xu2021applying}, and Curve Fit\cite{juliano2020nonlinear}. However, it has been shown~\cite{profile_RA} that regression is not well-suited for predicting saturation regions, while \ac{sl} models like \ac{ann} and \ac{knnr}, along with Interpolation, do not provide configuration trends with a monotonic rising function. In contrast, Curve Fit achieves high accuracy in predicting \ac{vnf} performance but is limited in multi-objective resource optimisation.
Moreover, \ac{sl} models explored for service-level \ac{vnf} profiling and placement may prove suitable under \textit{static} conditions~\cite{bunyakitanon2020auto},
however, they can significantly underperform under dynamic network conditions~\cite{2020:are3p} such as in contemporary networks.
Last, an important weakness of most \ac{ml} works is their approach to profiling as a \textit{single}-objective (i.e., single resource-type) optimisation problem, hence \textit{lacking realism} as most \acp{vnf} need more than one resource types, posing a non-linear impact of allocated resource amount combinations on resulting \acp{kpi}.

\subsection{Solution objectives}\label{sec:objectives}

The problem statement presented above establishes the context for the current solution effort, which revolves around four primary research objectives:

\textbf{Objective \#1:} The profiler should accommodate \textit{multiple} resource types and \acp{kpi}, and \textit{must} efficiently converge towards optimised \ac{vnf} configurations within \textit{practical timeframes}, 
despite the NP-hardness of the underlying optimisation problem.

\textbf{Objective \#2:}  Leverage \textit{online learning} \ac{ml} techniques to effectively adapt to the \textit{dynamics} of contemporary networks. 

\textbf{Objective \#3:} Investigate the impact of \textit{different} and \textit{pragmatic} \ac{vnf} types, with varying functionality features, on optimal resource allocation concerning specific \ac{kpi} targets.

\textbf{Objective \#4:} Conduct a comprehensive evaluation by comparing the proposed online learning solution against state of the art \ac{sl}-based \ac{vnf} profiler models.

In pursuit of these objectives, iOn-Profiler extends our prior work of~\cite{iprofiler} with an (i) in-depth analysis of the complex results obtained from an exhaustive search of resource-to-\ac{kpi} combinations, to (ii) gain a comprehensive understanding of the relevance of resource-to-\ac{kpi} and resource-to-\ac{vnf} type relationships, thus enabling to (iii) fine-tune the parameters of the online learning model in iOn-Profiler. Additionally, our extension involves considering (iv) a broader set of pragmatic \ac{vnf} types, encompassing different features, and exploring (v) multiple optimisation scenarios per \ac{vnf}. Lastly, the article presents a (vi) meticulous experimental evaluation of state of the art \ac{sl}-based \ac{vnf} profiler benchmark models, including \textit{\ac{rf}} and \textit{\ac{mlp}}. The presented research aims to advance the field of \ac{vnf} profiling and contribute valuable insights into enhancing the efficiency and performance of future network architectures.

Compared to the rest of the state of the art in \ac{ml}-based profiling (elaborated in Sec.\,\ref{sec:related-works}), iOn-profiler is designed to cover existing gaps (see Tab.\,\ref{table:related_works}). We go \textit{beyond single-objective} optimisation by utilising \ac{rl} to better fit \textit{real-world} applications while being \textit{adaptable} to network dynamics. 
We exploit carefully designed reward functions for the multi-objective optimisation of \ac{vcpu}, memory, and network \ac{lc} resource allocations that can achieve desirable \ac{vnf} \acp{kpi} targets such as the CPU utilisation, memory utilisation, latency and Optimum~\ac{or}.

To do so, our comprehensive study considers a wide spectrum of different scalarisation weights among \ac{vcpu}, memory and \ac{lc} objectives, which describe the \textit{Pareto front} of optimised resource-to-\ac{kpi} combinations that we wish to approach in 39 scenarios. 
The Pareto front is a concept 
representing the set of \textit{non-dominated} solutions\footnote{
    Non-dominated are best trade-off solutions, being \textit{impossible} to further one objective unless compromising at least one of the other optimisation objectives.
    }. 
When it comes to \ac{vnf} profiling the state of the art frequently ignores the Pareto front, posing a \textit{major research gap}. 
Even when considered, this refers primarily to \ac{sl} approaches tailored as ``static'' models trained for a given \ac{vnf} type, under specific conditions (e.g., network structure or traffic), and therefore \textit{cumbersome} or even impossible to generalise, if realistic at all for the highly agile and dynamic contemporary programmable networks.  

\begin{table*}[ht]
\caption{State of the art summary in intelligent VNF profiling. Compared to others, iOn-Profiler ``fills in'' all columns corresponding to research gaps.}
\label{table:related_works}
\begin{adjustbox}{width=\textwidth,center}
\begin{tabular}{|l|ccc|ccc|cccc|c|cccc|}
\hline
\multicolumn{1}{|c|}{Ref}                  & \multicolumn{3}{c|}{ML Techniques Used}                                                              & \multicolumn{3}{c|}{Considered Resources}     & \multicolumn{4}{c|}{Considered KPIs/Metrics}                 

& \multirow{2}{*}{\begin{tabular}[c]{@{}l@{}}Target\\ Platform\end{tabular}} & \multicolumn{4}{c|}{Predicting the optimum objectives/targets} 
\\ \cline{2-11} \cline{13-16}

& \multicolumn{1}{c|}{\begin{tabular}[c]{@{}c@{}}Supervised \\Learning\end{tabular} }      & \multicolumn{1}{c|}{\begin{tabular}[c]{@{}c@{}}Unsupervised \\Learning\end{tabular} }   & \multicolumn{1}{c|}{\begin{tabular}[c]{@{}c@{}}Reinforcement \\Learning\end{tabular} }  

& \multicolumn{1}{c|}{\begin{tabular}[c]{@{}c@{}}vCPU \\Cores\end{tabular}} & \multicolumn{1}{c|}{Memory} & \multicolumn{1}{c|}{\begin{tabular}[c]{@{}c@{}}Link \\Capacity\end{tabular}}

& \multicolumn{1}{c|}{\begin{tabular}[c]{@{}c@{}}vCPU \\Utilisation\end{tabular}} & \multicolumn{1}{c|}{\begin{tabular}[c]{@{}c@{}}Memory \\Utilisation\end{tabular}} & \multicolumn{1}{c|}{Latency} & \multicolumn{1}{c|}{Comments}  

& 

& \multicolumn{1}{c|}{\begin{tabular}[c]{@{}c@{}}vCPU \\Cores\end{tabular}} & \multicolumn{1}{c|}{Memory} & \multicolumn{1}{c|}{\begin{tabular}[c]{@{}c@{}}Link \\Capacity\end{tabular}} & \multicolumn{1}{c|}{Comments} 
\\ \hline

Orca~\cite{iglesias2017orca} & 
\begin{tabular}[c]{@{}c@{}}  \CheckmarkBold\\Regression\end{tabular} &  &  &  
 &  &  &
\CheckmarkBold &  & \CheckmarkBold & \begin{tabular}[c]{@{}c@{}} Throughput, \\Workload \end{tabular} &  
Docker &  
&  & & \begin{tabular}[c]{@{}c@{}} workload\end{tabular}
\\ \hline

Mestres et al.~\cite{mestres2018machine} &  
\begin{tabular}[c]{@{}c@{}}  \CheckmarkBold\\ANN\end{tabular} &  &  & 
\CheckmarkBold &  &  & 
&  &  & \begin{tabular}[c]{@{}c@{}} Throughput, \\Response time \end{tabular} &  
\begin{tabular}[c]{@{}c@{}} Non-specified\\Hypervisor\end{tabular}&  
\CheckmarkBold & &  & 
\\ \hline

WRVS2~\cite{Manuel2018} &   
\begin{tabular}[c]{@{}c@{}}  \CheckmarkBold\\Regression\end{tabular} &  &   &    
\CheckmarkBold &  &  &
&  &  & Throughput &  
SOTANA &     
&  & & N/A
\\ \hline

\begin{tabular}[c]{@{}c@{}}Van et al.~\cite{profile_RA}\end{tabular} & 
\begin{tabular}[c]{@{}c@{}}  \CheckmarkBold\\Regression, kNN, \\Interpolation, \\ANN, Curve Fit\end{tabular}  & & & 
\CheckmarkBold & & \CheckmarkBold & 
\CheckmarkBold & & & Packet loss &  
Docker &  
\CheckmarkBold &  &  & \begin{tabular}[c]{@{}c@{}}  Packet rate, \\Response time\end{tabular} 
\\ \hline

z-torch~\cite{ztorch} & 
& \CheckmarkBold  &  &  
\CheckmarkBold & \CheckmarkBold & \CheckmarkBold &
\CheckmarkBold & \CheckmarkBold & \CheckmarkBold &  &  
OpenEPC &  
&  &  & \begin{tabular}[c]{@{}c@{}}  Opt. VNF-\\Placement\end{tabular}
\\ \hline
 
{\begin{tabular}[c]{@{}c@{}}Van et al.~\cite{van2020vnf} \end{tabular}}&    
\begin{tabular}[c]{@{}c@{}}  \CheckmarkBold\\ANN, GP,\\ Interpolation, \\Regression\end{tabular}  &  &  &  
\CheckmarkBold &  &  &
\CheckmarkBold &  &  & \begin{tabular}[c]{@{}c@{}}  Response time,\\Packet rate,\\ Workload\end{tabular} &  
Docker&  
& &  & N/A  
\\ \hline

NFV-Inspector~\cite{khan2018nfv} & 
\begin{tabular}[c]{@{}c@{}}  \CheckmarkBold\\Decision Tree\end{tabular}&  &   & 
\CheckmarkBold & \CheckmarkBold & &
&  &  & Arrival rate &  
NFV Inspector & \CheckmarkBold 
& \CheckmarkBold&  & \begin{tabular}[c]{@{}c@{}}  Disk,\\Bandwidth\end{tabular} 
\\ \hline

NAP~\cite{NAP} & 
\begin{tabular}[c]{@{}c@{}}  \CheckmarkBold\\MIMO-GRNN,\\
RF, MLP\end{tabular} &  &  & 
\CheckmarkBold & \CheckmarkBold & \CheckmarkBold &
\CheckmarkBold & \CheckmarkBold & \CheckmarkBold & Input rate &  
Docker &  
\CheckmarkBold & \CheckmarkBold & \CheckmarkBold & Optimum Input rate 
\\ \hline
\Revised 
RAVIN~\cite{ravin} & 
\begin{tabular}[c]{@{}c@{}} N/A (not ML) - BBFD heuristic\end{tabular} &  &  & 
LLC & MB  &  &
 &  &  & Throughput &  
Custom C++ platform &  
\CheckmarkBold & \CheckmarkBold &  & number of Servers
\\ \hline
\Revised 
PDPA~\cite{vnf-prof} & 
\begin{tabular}[c]{@{}c@{}}  \CheckmarkBold\\RF, KNN\\
DT, MLP, LR\end{tabular} &  &  & 
\CheckmarkBold & \CheckmarkBold & \CheckmarkBold &
 &  & \CheckmarkBold & Packet loss, Input rate &  
OSM &  
\CheckmarkBold & \CheckmarkBold & \CheckmarkBold &  
\\ \hline
\Revised 
Ru et al.~\cite{perfo-bottleneck} & 
\begin{tabular}[c]{@{}c@{}} \CheckmarkBold\\Classifier\end{tabular} &  &  & 
\CheckmarkBold & \CheckmarkBold &  &
 &  & \CheckmarkBold & Packet level performance, Throughput &  
Cisco VPP &  
 &  &  & performaqnce perturbation 
\\ \hline

\begin{tabular}[c]{@{}c@{}}\textbf{iOn-Profiler}\end{tabular}  &  
\begin{tabular}[c]{@{}c@{}}  \CheckmarkBold\\RF, MLP\end{tabular} &  & \begin{tabular}[c]{@{}c@{}}  \CheckmarkBold\\Q-Learning\end{tabular} &  
\CheckmarkBold & \CheckmarkBold  & \CheckmarkBold  &
\CheckmarkBold & \CheckmarkBold & \CheckmarkBold & Output rate & 
\begin{tabular}[c]{@{}c@{}} MANO\\ (OSM) \end{tabular} & 
\CheckmarkBold & \CheckmarkBold & \CheckmarkBold & \begin{tabular}[c]{@{}c@{}}  Optimum\\Output rate 
\end{tabular} 
\\ \hline

\end{tabular}
\end{adjustbox}
\end{table*}

\subsection{State of the art}
\label{sec:related-works}

The state of the art discussed below is summarised and compared in Tab.\,\ref{table:related_works}.
First off, various important works~\cite{profile_chains, Manuel2018, mestres2018machine} have explored \textit{offline} profiling for \ac{vnf} \acp{sfc} with a focus on different resources. 
Regarding optimal \ac{vnf} placement and profiling, RAVIN~\cite{ravin} introduces a resource-aware algorithm based on the Balanced Best Fit Decreasing (BBFD) heuristic algorithm. It enforces performance \acp{sla} for multi-tenant \ac{nfv} servers while balancing resource use, aiming to minimize server count, guarantee performance, and improve resource utilization, including the processor's Last Level Cache and Memory Bandwidth (MB).
    However, extensive offline profiling by exploring all possible \ac{vnf} configurations such as in the aforementioned works is \textit{time-consuming}, 
    leading to the development of models that focus on limiting the profiling time such as in~\cite{Manuel2018}. 
    Nonetheless, and unlike our current effort in iOn-Profiler, endeavours like~\cite{Manuel2018} do not encompass the concurrent consideration of pivotal \acp{kpi} such as \ac{vcpu} utilization, memory utilization, latency, throughput, and packet loss.
    In another study by~\cite{perfo-bottleneck}, researchers address the challenges of diagnosing \ac{nfv} performance and introduce a metric referred to 
    as the \textit{Coefficient of Interference}. 
    This metric quantifies the variations observed in latency measurements on a per-packet basis when performance diagnosis is applied against cases in which it is not employed.
    Last, other works, such as ORCA~\cite{iglesias2017orca} and z-TORCH~\cite{ztorch}, have streamlined the profiling process for data collection and optimal \ac{vnf} placement. 
    However, these approaches may not consider optimal \acp{kpi} and pre-defined resource configurations.
Last, other notable contributions in the literature include the NFV-Inspector~\cite{khan2018nfv}, an automated profiling and analysis platform, and the work of~\cite{van2020vnf} 
utilising \ac{ml} techniques such as Interpolation, Gaussian Process, \ac{ann}, and Linear Regression for predicting \ac{vnf} performance.

Regarding our own contributions to the field of \ac{vnf} profiling,
    the \acf{nap} method~\cite{NAP} focuses on offline autonomous profiling by 
    identifying the initial optimal resource configuration for each standalone \ac{vnf} based on a 
    weighted resource configuration selection approach. 
    Furthermore, our most recent work of~\cite{vnf-prof} introduces a novel autonomous \textit{temporal} profiling technique, 
    examining \ac{vnf} behaviour across performance and resource utilisation aspects. 
    The proposed technique automates the profiling processes, encompassing diverse resource types like computation, memory, and network resources, 
    to yield deeper insight into \acp{vnf} resource-performance correlations. 
    Finally, further to our prior works and, particularly, \ac{nap}\cite{NAP}, the current method in \textit{iOn-}Profiler 
    spans an offline training and an online learning phase that enables adopting network dynamics at deployment time, both grounded in \ac{rl}.
    Moreover, \textit{iOn-}Profiler deploys \acp{vnf} on the established \ac{mano} platform, 
    namely \ac{osm} \cite{yilma2020benchmarking}, and suggests a \textit{multi-objective} \acp{vnf} profiling strategy also grounded in \ac{rl}. 
    Fitting \ac{rl} profiling agents into a more complex \ac{rl} model-based \textit{orchestration} scheme is possible with a \textit{hierarchical} RL structures~\cite{helicon} allowing to place VNFs to nodes and allocating resources there leveraging multiple sources of information spanning from VNF profiles, system-wide resource usage information~\cite{helicon} and even service consumers mobility\cite{uniyal21}.

In conclusion, using \ac{ml} for \ac{vnf} profiling has been extensively studied in various domains. These studies collectively demonstrate that \ac{ml} significantly enhances the accuracy and other qualitative features of \ac{vnf} profiling compared to traditional methods. As the field of intelligent \ac{vnf} profiling continues to evolve, further advancements in \ac{ml}-based approaches hold promise for improving network performance and resource optimisation. 
In this context, this paper stresses the advantages of \ac{rl} compared to other intelligent solutions, 
covering all features but that of \ac{sl} model used in Tab.\,\ref{table:related_works}.

\begin{figure}[h]
    \centering
    \includegraphics[width=0.6\linewidth]{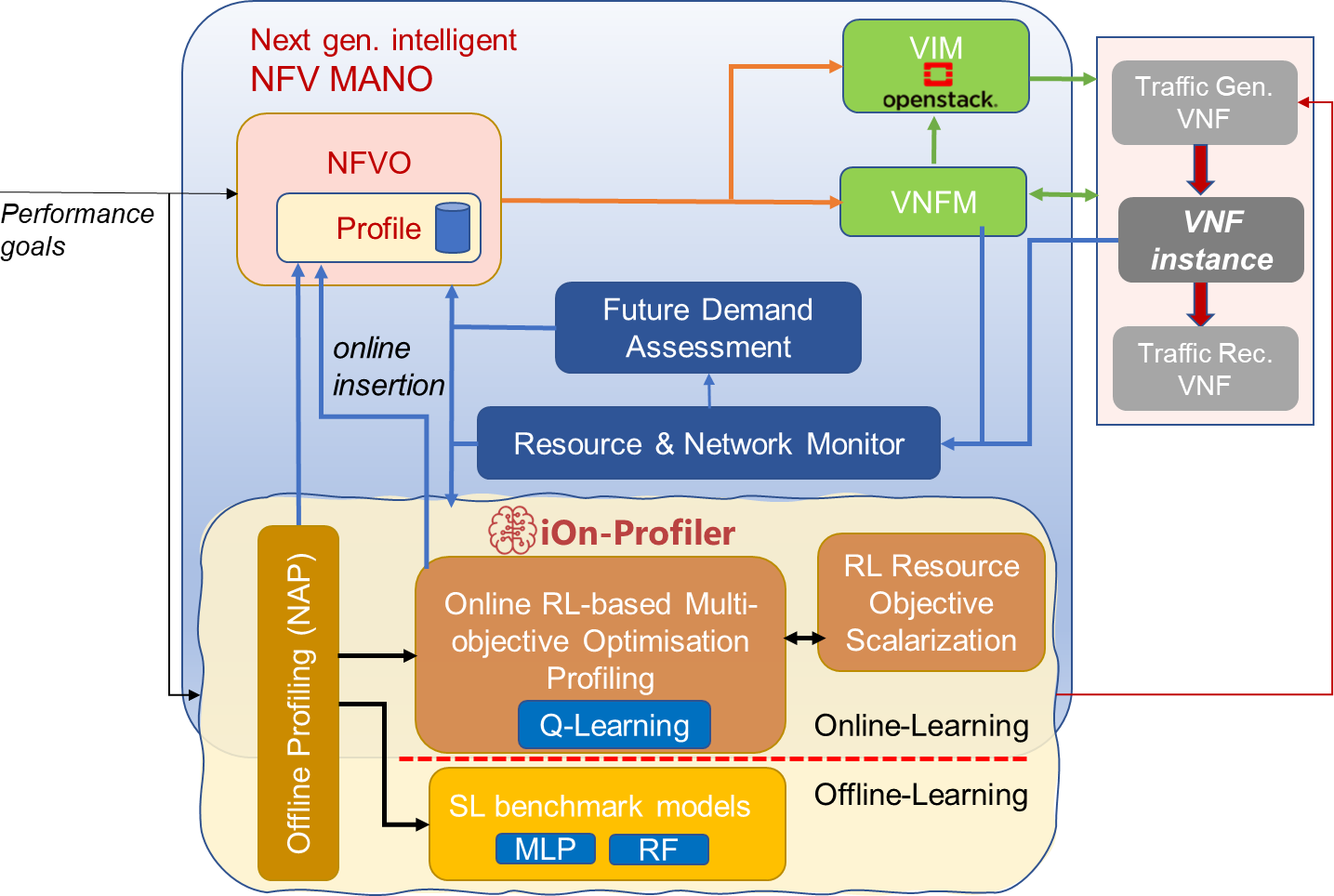}
    \caption{iOn-Profiler and MANO integration diagram.} 
    \label{fig:interaction-v2}
\end{figure}

\section{iOn-Profiler model design}
\label{sec:profiler-design}

\subsection{Integration into the next generation NFV MANO}
\label{sec:mano-vs-profiler}

Figure~\ref{fig:interaction-v2} illustrates the interaction between our proposed iOn-Profiler, \ac{nfvo}, \ac{vim}, and monitoring tools to provide an intelligent and autonomous  \ac{nfv} \ac{mano} system. The diagram not only shows interaction but also demonstrates the integration of the iOn-Profiler into next-generation intelligent \ac{nfv} \ac{mano}. Through online profiling, the configuration of resources is selected and dynamically updates existing virtual network function (\ac{vnf}) descriptors. As a result, the \ac{mano} system deploys a network slice with the newly defined resources. In Fig.~\ref{fig:interaction-v2}, we outline iOn-Profiler's architecture.

\subsubsection{Offline Profiling}\label{sec:offline-prof}

Given a series of resource availabilities and a number of \ac{kpi} targets, iOn-Profiler employs the NAP method~\cite{NAP} to select a baseline resource configuration. Resources\footnote{
    Resource types that can be either present or not such as 
    smart Network Interface Cards mapped via single root I/O virtualization can be captured via profiling two corresponding VNF distinct variants.
} and \acp{kpi} types can be arbitrary, provided they are described by a value in a totally ordered bounded set with at least 3 elements (see 
Sec.~\ref{sec:ExpParamsConf}, Tables~\ref{table:config} and~\ref{table:KPItargets}).
NAP is based on the concept of optimal \ac{ir} and \ac{or} and can be divided in 3 stages. 
The optimal IR is the maximum IR (in packets per second) associated with a specific resource configuration for which the system under test still respects all KPI targets, and the optimal \ac{or} is the \ac{or} associated with it.

In stage 1, NAP employs exponential ramp-up and binary search to find the optimal \ac{ir} for each resource’s upper/lower bounds while other resources remain at their median. Using these values, weights are calculated to measure resource influence on performance.
In stage 2, NAP uses weighted random selection for applying resource configurations, measuring \ac{ir}, \ac{or}, and \acp{kpi}. In stage 3, NAP trains a model to estimate minimum resource allocation based on \ac{ir} and KPI targets.
The method uses the NFVO and VIM to deploy the VNFs, a traffic generator and a monitoring probe at each step. Traffic generators can be employed to overcome (a) a possible lack of available real traffic datasets 
and (b) the need for fine-tuning the IR as required by the algorithm. 
We refer to the above as \textit{offline profiling} as it needs a dataset implying the control of \ac{ir} for generating arbitrary network traffic conditions, and training before the model can be used 
in production.

\subsubsection{Online Multi-Objective Optimisation Profiling}\label{sec:env-adapt-prof}

{\Revised 
After deploying the VNF with baseline resources, the iOn-Profiler employs Q-Learning (see Sec.~\ref{sec:rl-algorithm}) to address possible discrepancies after moving from the staging environment where the \textit{Offline Profiling} regression model is trained, to an \textit{online dynamic} environment. Possible disparities are recognised when the target KPI thresholds are breached, prompting the resetting of the exploration rate and other learning parameters (see Sec.\,\ref{sec:env-adapt-prof}). This continuous optimisation tries to minimise resource usage without violating KPI targets and to improve allocation accuracy. 
Therefore, the optimisation objectives need to match the same set of resource types selected for the offline profiling, subject to the same restrictions.
Each action uses the NFVO and VIM APIs to scale in/out the VNF instance and the exposed monitoring capabilities.
Last, the term \textit{online profiling} is due to the profiler (i) observing only existing network traffic without control over \ac{ir}, and (ii) being used in a production environment.
}

\begin{algorithm}[h]
\caption{Multi-objective Q-Learning adaptation.}\label{alg:scale_multiobj}
          \SetAlgoLined
          \SetKwInput{KwData}{Input}
          \KwData{
              learning rate $\alpha = 0.1$,
              discount factor $\gamma = 0.99$,
              the best steepness coefficient value ($\beta$) for each reward function,
              {\Revised
              maximum number of steps ($N$),
              convergence check threshold ($\varepsilon$),
              number of steps for convergence check ($N_\varepsilon$)
              }%
              .
          }
        
        %
        {\Revised
        \For{$each\: objective\: o$\label{line:con_Qtable}}{
           Initialise $Q_o(s, a)$ as an empty \:Q-table.\label{line:proc_create_Qtable}
        }
        }
        
        \For{$each\: episode\: t$\label{line:loop_start_ep}}{
            Initialise present state $s$ vector\label{line:set_Pstate}\; 
            {\Revised
            Initialise circular buffer $\Delta s$ with $N_\varepsilon$ slots\;
            }
        
        \For{\Revised $n \gets 0$ to $N$}
        {
        \eIf{$\rand(0,1) <$  $\epsilon$ -greedy \: \label{line:cond_epsilon}}{
            Action $a \gets \rand \: A(s)$\label{line:get_randAction}\;
          }{
          {\Revised
           
          Action $a \gets$ Call Algorithm \ref{alg:scal_greedy} for state $s$; \label{line:con_epsilon_end}}
          }
          Take \:action \:$a$ \:and \:observe \:the \:next \:state ${s}'$\label{line:proc_takeAction}
          
          Calculate \:reward ($\boldsymbol{R}_{o}$) of each resource in ${s}'$ through equation (\ref{eq:rew_o})\label{line:proc_reward}. 
        
         Call \:Algorithm \:\ref{alg:scal_greedy} \:to \:find \:${a}'$ based on ${s}'$ that gives the maximum Scalarised Q-value\label{line:proc_callAlg1}
        
            \For{$each\: objective\: o$\label{line:loop_for_bellman}}{
                $Q_{o}(s,a) \gets Q_{o}(s,a)+\alpha \left ( \boldsymbol{R}_{o} +\boldsymbol{\gamma}Q_{o}({s}',{a}')-Q_{o}(s,a)\right )$\label{line:get_bellman};     
            }
        Ask \:the \:NFVO \:to \:scale \:in \:or \:out the \ac{vnf} \:based \:on \:${s}'$\label{line:proc_scaling}
        
        {\Revised Find the \ac{or}} and record the corresponding state\label{line:proc_profiling}
        
        {\Revised
        Insert $\lVert s - s' \rVert$ in $\Delta s$;
        }
        
        $s \gets s'$\;\label{line:get_Pstate}
        
        {\Revised
        \If{$n > N_\varepsilon$ \textbf{and} $\max \Delta s < \varepsilon$}{
            \text{The algorithm has converged;}\\
            \textbf{stop}
        }
        }
        }
        }
\end{algorithm}

\subsection{Multi-objective reinforcement learning model adaptation}\label{sec:rl-algorithm}
    {\Revised Algorithm~\ref{alg:scale_multiobj} describes} our multi-objective \ac{rl} approach to optimising resource allocation for a given type of \ac{vnf}. This approach is aimed at addressing a Markov decision process
    {\Revised
    by dynamically constructing Q-tables ($Q_o$) for each optimisation objective ($o$) that stores the estimated discounted sum of future rewards for each possible action ($a$) at a given state ($s$).
    The Q-tables gradually converge by exploring the action space and performing updates based on the recursive Bellman equation (shown in line~\ref{line:get_bellman})
    Our model considers the following definitions for state ($a$), action ($a$) and reward ($R_o$):
    }

\subsubsection{State}
    A vector that encompasses allocated resources (e.g., \ac{vcpu} cores number) 
    in addition to the measured \acp{kpi} (e.g., \ac{vcpu} utilisation) 
    and \ac{or}. 

{\Revised 
\subsubsection{Action}
    The set of feasible actions encompasses increasing, decreasing, or preserving resource assignments. 
    These actions induce shifts between various states of allocation
    (e.g., incrementing/decrementing the number of vCPU cores).
    In terms of action choice, we employ a scalarized $\epsilon$-greedy algorithm, which facilitates the selection of actions that optimise individual rewards for each resource category by selecting the action with the highest reward 
    with probability $1-\epsilon$.    
}
%
%
\begin{algorithm}[h]
\caption{Scalarised Greedy Action Selection.}
\label{alg:scal_greedy}
  \SetAlgoLined
  \SetKwInput{KwData}{Input}
  \KwData{
      $w_o \gets \text{The weight of each objective}$,
      {\Revised $s \gets \text{observed state}$}.
  }
$SQlist \gets \{\}$\label{line:set_SQlist}\;
\For{$each\: action\: a\in A$\label{line:loop_for_action}}{
    $v \gets \hat{Q}(s,a)=\left \{ Q_1(s,a),Q_2(s,a),...,Q_m(s,a) \right \}$\label{line:proc_Qvalue}\;
    $\widehat{SQ}_{linear}(s,a) \gets f(v,w)$\label{line:proc_scalarize}\;
    Append $\widehat{SQ}_{linear}(s,a)$ to SQlist\label{line:proc_append}\;
}
\textbf{return} $\underset{a'}{\argmax}\;SQlist\label{line:proc_ScalizedAction}$
\end{algorithm}
%
%
\subsubsection{Reward Function}
    To find the reward function for each \ac{vnf} type, we consider and optimise the parameters of the following reward function model.
For each resource type (i.e., objective), we use the \textit{zedoid} (i.e., a reverse sigmoid) general formula of $f(x) = \frac{1}{1+e^{x}}.$ Zedoid function allows to adaptively/gradually yield reward values that \textit{decay with increased resource} allocations and vice versa. Therefore, the rewards promote a more cost-efficient use of resources. 

\begin{figure}[h]
    \centering
    \includegraphics[width=0.35\columnwidth]{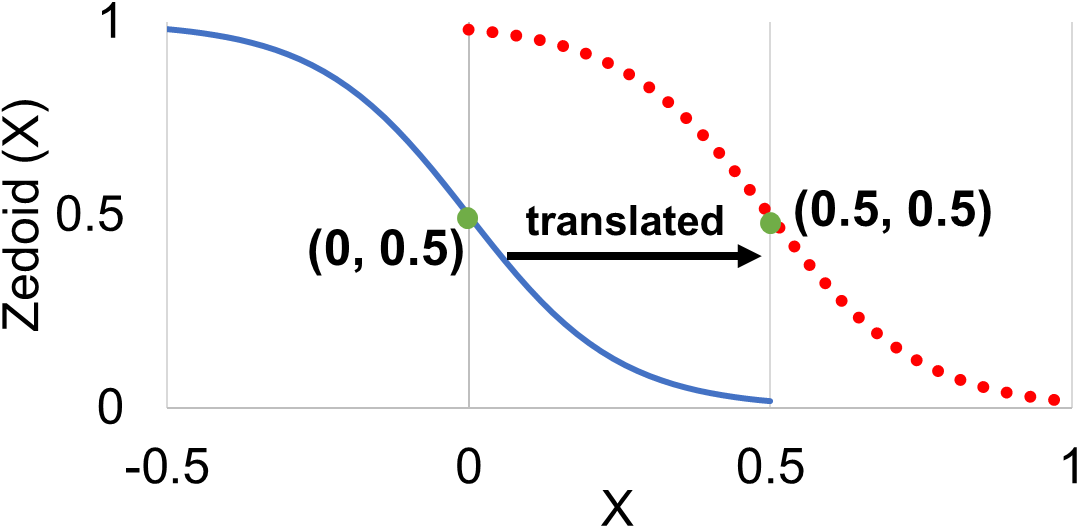}
    \caption{Translation of \textit{zedoid} function by 0.5 units to yield rewards only for positive resource allocation over the x-axis.}
    \label{fig:zedoid}
\end{figure}

We adopt an appropriately parametrised (discussed in Sec.~\ref{sec:eval:Qparams}) version of the zedoid function depicted in Fig.~\ref{fig:zedoid}. The adopted zedoid is shifted by 0.5 units. This is a desired transposition of the zedoid curve so that reward values reflect meaningful (i.e., positive) resource allocations over the x-axis. 
It is worth noting that in Fig.~\ref{fig:zedoid}, the blue solid graph curve and the translated red dotted curve follow the $\frac{1}{1+e^{8x}}$ and $\frac{1}{1+e^{8(x-0.5)}}$ formulas, respectively.  
%
{\Revised
We also impose a penalty for constraint violation (including KPI targets)
by mapping the computed value to 0.
}
The general formula of the adopted reward function for each resource type is defined in \,(\ref{eq:rew_o}):
{\Revised
\begin{equation}
\centering
    \label{eq:rew_o}
    R_{o} = \begin{cases}
    \dfrac{1}{1+e^{\beta\left ( \hat{o} - 0.5\right )}}, & \text{constraints satisfied} \\
    0, & \text{otherwise}        
    \end{cases}
\end{equation}
}
where $\hat{o}$ is allocated resource, i.e., the number of allocated \ac{vcpu} cores, the amount of allocated memory or \ac{lc}; 
and $\beta$ is the steepness coefficient of the resource reward function that defines a desired curve steepness best fitting a resource type's adaptability to allocation changes.
We return and optimise the selection of $\beta$ in Sec.~\ref{sec:eval:Qparams}.
%
We extend the scalarisation function from single to multiple objective calculations, as in Algorithm~\ref{alg:scal_greedy}. For each action at line \ref{line:loop_for_action} - \ref{line:proc_Qvalue}, 
Q values from all objectives are put in a vector as (\ref{eq:qvalues}). Note that $m$ refers to each optimisation objective.
{\Revised  This vector and a weight vector $w = (w_1, w_2, \ldots, w_m)$ are applied to the scalarisation function $f(v,w)$ to calculate the scalarised Q-value (SQ) according to (\ref{eq:sq_linear})}.
The sum of all weights must be 1. 
%
At line \ref{line:proc_append}, {\Revised SQ is} appended to the $SQlist$. Finally, at line \ref{line:proc_ScalizedAction} the algorithm returns the action ${a}'$ corresponding to the highest SQ.
    \begin{equation}
        \label{eq:qvalues}
        \hat{Q}(s,a)=\left \{ Q_1(s,a),Q_2(s,a),...,Q_m(s,a) \right \}
    \end{equation}
    \begin{equation}
    \label{eq:sq_linear}
    \widehat{SQ}_{linear}(s,a){\Revised \, = f(v,w) =} \sum_{o=1}^{m} w_{o}\,Q_o(s,a), \text{where} \sum_{o=1}^{m} w_{o}= 1
    \end{equation}
%

{\Revised  
    \subsection{Solution complexity and practical costs}
    \label{sec:complexity}

     The execution (\textit{time} and \textit{memory}) complexity of iOn-Profiler is defined by the interaction between actions and the state space of the underlying Q-learning process.
     Each state includes the allocated resource values, \ac{kpi} thresholds, parameters referring to the input passed to the \ac{vnf} (e.g., input requests traffic), and last, the observed \ac{kpi} measurements.    
     As such, the state space complexity is defined by the count of
        (a) the resource types considered,
        (b) the input types, and 
        (c) the targeted \acp{kpi}. 
     Given the former, 
     the asymptotic execution complexity is also a function of 
          (i) the \textit{granularity} of possible resource assignment levels per resource; 
          (ii) the number of resources; 
          (iii) 
          the measurement granularity per \acp{kpi}; 
          and last the (iv) possible input levels per each input type. 
          
               

Further to the problem definition (Sec.\,\ref{sec:probState}), 
let $\lambda =|I|$ be the number of resource types considered and $\kappa= |K|$ the count of the different \acp{kpi} targeted.
Let $\Psi$ be the set of \ac{vnf} input categories, with $\zeta = |\Psi|$.  
Also, let each $i$ be assigned values in $\{\iota_1, \iota_2, ... \iota_{\rho}\}$. Note that the granularity of the latter counts $\rho$ feasible resource level allocation options for each $i$.  
For coherence, we classify measurements for each $k$ into the immediately preceding class within set $T_k$, thus counting $\omega$ measurement options (Recall $T_k$ from Sec.\,\ref{sec:probState}). 
Finally, let set $U_\psi = \{\upsilon_1, \upsilon_2, ... \upsilon_{\eta}\}$ define a partition of possible \ac{vnf} input levels per type $\psi \in \Psi$, thus counting $\eta$ levels.

    The above gives an \textit{an asymptotic upper bound} for the space complexity and time needed to explore the whole space of $O(\rho^\lambda) \times O(\omega^\kappa) \times O(\eta^\zeta)$, which falls within the \textit{exponential} complexity class $O(c^n) | c > 1$.
     %
     \textit{Practical implementations} of Q-learning solutions set thresholds for action steps, 
     hence reducing memory and time costs significantly. Another aspect of time costs 
     refers to the adopted learning rate in \ac{rl}. 
     
     The demonstrative implementation setup considered in the current paper assumes 
        \textit{three} types of resources;
        \textit{one} input type: input traffic to the \ac{vnf}; 
        and \textit{four} target \acp{kpi} with one threshold target each (see Tab.\,\ref{table:KPItargets}). 
     As such, the state space is represented by a \textit{vector} of \textit{nine} elements, including the allocated \ac{vcpu} cores, memory, and output \ac{lc}, the four \ac{kpi} measurements, and input traffic value, and the computed scalarised $Q$ value. 
     Given the adopted resource configuration values in Tab.\,\ref{table:config} and allocation steps in Sec.\,\ref{sec:ExpParamsConf}, the implied memory needs include nine 16-bit float numbers (i.e., 144 bits) per 
     state\footnote{
        Note, that the previous numbers can be significantly compressed by using bitmaps of deltas rather than 16-bit float or short numbers. Caching recently computed $Q$ values can also save a lot of exploration/exploitation.
    }.   
    Given the resource levels assumed in Sec.\,\ref{sec:ExpParamsConf}, and information in Tab.\,\ref{table:config} and Tab.\,\ref{table:KPItargets}, 
    there are $6 \times 6 \times 8 = 288$ resource level combinations, and $2^4 = 16$ \ac{kpi} alignment/violation combinations.  
    As a result, there are $288 \times 16 = 4608$ states in the state space implying an 81\,KB memory need. 
    This is marginally lower than typical first-level CPU data cache sizes (e.g., 16-128KB), thus allowing us to benefit from fast computations. 
    %
    Besides the memory cost, the mean time cost of each training episode in our experiments is 738 steps, defined by the convergence of the $Q$ value.

    
}

\section{Experimental setup}
\label{sec:eval-setup}

Fig.~\ref{fig:setup} depicts our profiling experimental setup assuming \snort or \ac{vfw} as the \ac{vnf} instance. 
It shows the connection between the profiled \ac{vnf} on the one hand, and the traffic generator and server end-point machines
on the other. The two end-point machines have two \ac{vcpu} cores, 2\,GB of memory and 10\,GB of storage.
{\Revised
For simplicity, we employ iPerf as a traffic generator with UDP packets, noting that active data collection is more suitable for \ac{rl} than static datasets.
}

\begin{figure}[h]
    \centering
    \includegraphics[width=.5\linewidth]{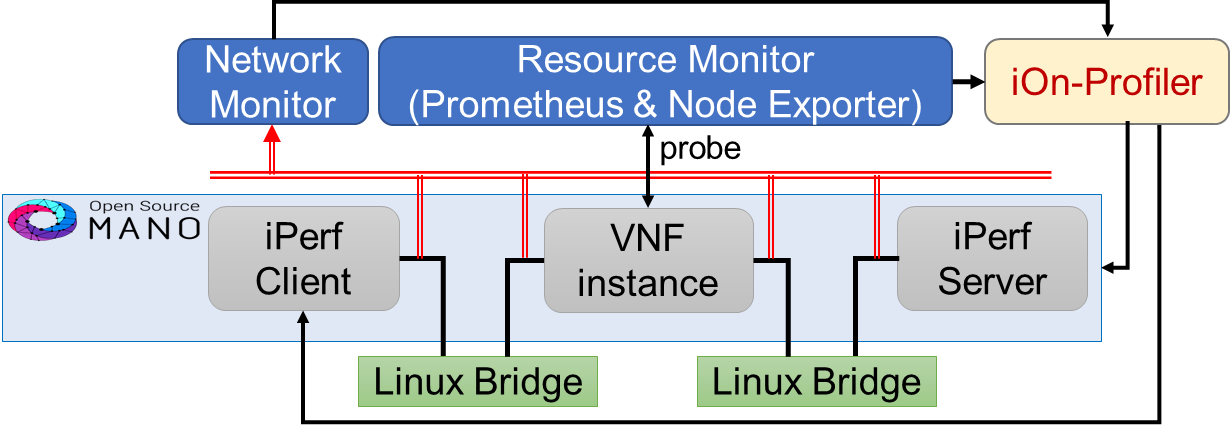}
    \caption{Experiment setup}
    \label{fig:setup}
\end{figure}


We employ the Prometheus and Node exporter monitoring tools to gather the following metrics: \ac{vcpu} utilisation, memory utilisation, and ingress and egress traffic rates to and from the \ac{vnf}, respectively.
Additionally, we calculated the mean \ac{rtt} using the ping utility.
In addition, the duration for the offline profiling was set to 48 hours for each \ac{vnf} model. The software tools and frameworks used in this study are outlined in Tab.\,\ref{table:software_details}.
%
\begin{table}[h]
    \caption{Software frameworks and tools.}
    \label{table:software_details}
    \centering
        \begin{tabular}{lll}
        \hline
        Software type & 
        Functionality provided & 
        Version \\ 
        \hline
        OSM & Orchestrate \& run \vnf instances & 8.0.4-1 \\
        \snort & Inline \& Passive mode & 2.9.17 \\
        \iperf3 & Generate traffic \& measure the bandwidth & 3.1.3 \\
        Ubuntu & The OS on each VM & 20.04 \\
        Prometheus & Monitor the performance metrics & 2.25.0 \\
        Node exporter & Gather information from Linux services & 1.1.2 \\ 
        \hline
        \end{tabular}
\end{table}

\subsection{VNF type scenarios}
\label{sec:eval-vnfs}

We evaluate the performance of our proposed method using three different types of \acp{vnf} as our experimental scenarios. These \acp{vnf} cover a range of scenarios and demonstrate varying sensitivities to different resources. For example, the performance of the copying \acp{vnf} may be more impacted by memory utilisation, while the intercepting \acp{vnf} may be more impacted by \ac{vcpu} utilisation.

\begin{enumerate}

    \item \textbf{\snort (Inline mode):}
    The \snort \ac{vnf} operates as a traffic gateway between network segments and inspects all incoming packets before forwarding them to the destination. This mode slows down traffic transmission and may block suspicious packets.
    
    \item \textbf{\snort (Passive mode):}
    The Passive mode \snort \ac{vnf} operates outside of the direct traffic path and copies incoming traffic to detect suspicious activity. This mode raises a different set of resource needs compared to the Inline mode, as shown by our evaluation results.
    
    \item \textbf{Virtual Firewall (vFW):}
    Allows packets to pass only through specified ports towards the destination server.

\end{enumerate}

{\Revised 
\subsection{Resource and KPI targets configuration}
\label{sec:ExpParamsConf}
}

{\Revised
We consider vCPU cores, memory and \ac{lc} as our profiled resources.
Tab.\,\ref{table:config} shows the upper and lower bounds for their
configuration values, chosen in accordance with our experimental environment and the specs of the considered \ac{vnf} types. The iPerf traffic generator client transmits UDP packets with an initial traffic rate of 50\,Mbps to the destination iPerf server. 
The traffic rate gets gradually increased, and the assumed \ac{kpi} thresholds are specified in Tab.\,\ref{table:KPItargets}.}
%
\begin{table}[htbp]
\centering
\caption{Resource configuration.} 
\label{table:config}
\begin{tabular}[t]{lc>{\color{black}}c}
\toprule
\textcolor{black}Configuration
& \textcolor{black}{values} 
\\
\midrule
Minimum and Maximum value of Resources: & \\
\quad number of vCPU cores & 0.6 - 1.8\\
& (30\% to 90\% of 2 cores) \\
\quad Memory (MB) & 1000 - 1600\\
\quad Link Capacity (Mbps) & 400 - 800\\
\bottomrule
\end{tabular}
\end{table}%

%
\begin{table}[htbp]
  \centering
  \caption{KPI targets that the \ac{vnf} under profiling should meet.}
  \label{table:KPItargets}
  \begin{tabular}[t]{lc>{\color{black}}c}
    \toprule
    \textcolor{black}KPI targets
    & \textcolor{black}{values} 
    \\
    \midrule
    \ac{vcpu} Utilisation (\%) & 95$\pm$5\\
    Memory Utilisation (\%) &$\leqslant$ 98\\ 
    Latency (ms) &2.5$\pm$5\\
    \bottomrule
  \end{tabular}
\end{table}
%
{\Revised 
\subsection{State transition actions and training episodes}
\label{sec:ExpStateParamsConf}

In terms of actions, 
\ac{vcpu} is increased/decreased by 0.2\,cores, memory by 100\,MB, and \ac{lc} by 50\,Mbps, which aligns quantisation of state space and implementation necessities after the complexity analysis in Sec.\,\ref{sec:complexity}. For the scalarised $\epsilon$-greedy algorithm, we adopt a decay factor $\epsilon=0.9999$, with minimum exploration rate 0.1, learning rate $\alpha=0.1$, and discount factor $\gamma=0.99$.
Training is organised in episodes encompassing action steps until either a maximum number of steps is reached or the minimum resources are found. Given this setup, a total of 2000 episodes was assumed, encompassing a mean number of 738 steps per episode.
}

\subsection{Rewards configuration}
\label{sec:eval:Qparams}
We conducted an experiment-based parameter tuning of our reward functions to optimally adjust parameters, such as the steepness coefficient $\beta$, to the unique requirements and characteristics of the three different \ac{vnf} types and to the impact of the three different resource types on profiling performance. This resulted in 9 individual reward parameterisations. To achieve this, we analysed each resource type for each \ac{vnf} type in isolation. This involved using a fixed resource allocation value for the other two resource types to speed up the process. The mean values of the other two resource types, which yield optimal allocation of the investigated resource type in a controlled environment (i.e. minimum resource usage for maximum \ac{or}), were used as the fixed values. 

\subsubsection{\snort (Inline mode)}

\begin{figure*}[h]
    \centering 
    \begin{subfigure}{0.45\columnwidth}
        \includegraphics[width=\textwidth]{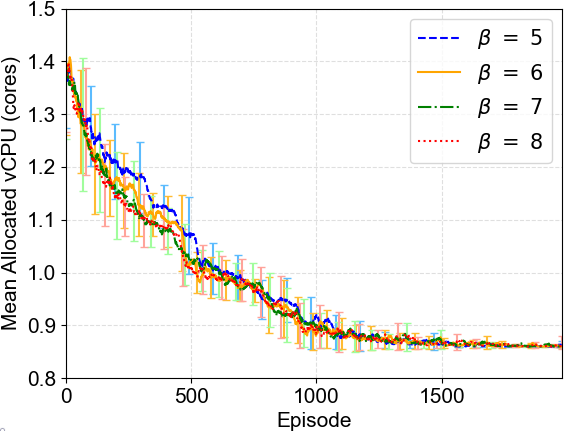}
        \caption{}
        \label{fig:lin_exp_inline_a}
    \end{subfigure}
    \begin{subfigure}{0.45\columnwidth}
        \includegraphics[width=\textwidth]{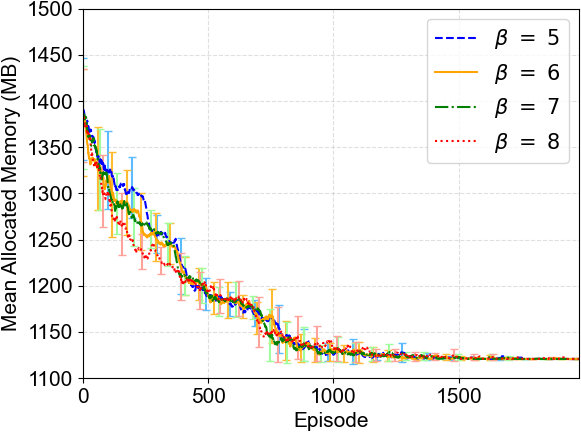}
        \caption{}
        \label{fig:lin_exp_inline_b}
    \end{subfigure}
%
    \begin{subfigure}{0.45\columnwidth}
        \includegraphics[width=\textwidth]{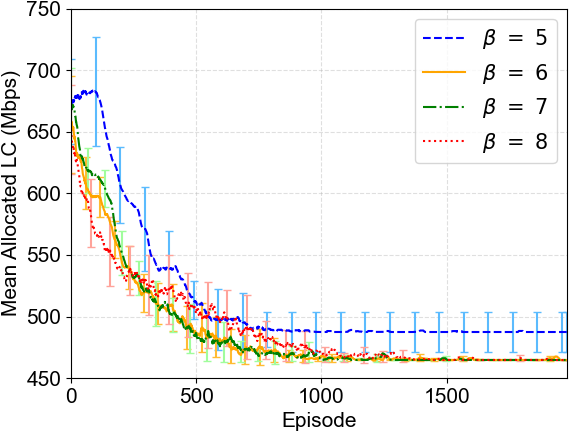}
        \caption{}
        \label{fig:lin_exp_inline_c}
    \end{subfigure} 
    \caption{Configuring reward function parameters for \snort with Inline mode.}
    \label{fig:lin_exp_inline}
\end{figure*}

\begin{itemize}
    \item 
        \textbf{vCPU ($R_{cpu}; \beta = 8$).} 
        Regarding \ac{vcpu}, Fig.~\ref{fig:lin_exp_inline}(a) shows how the model learns to adapt \ac{vcpu} values from 0.6 to 1.8 cores, assuming fixed mean memory and link capacities equal to 1300\,MB and 600\,Mbps, respectively. 
        Reward $R_{cpu}$ (for all different $\beta$ configurations) exhibits a similar performance, and convergence point after episode 1000, resulting in 0.88 \ac{vcpu} cores. 
        However, the best reward function based on the smallest confidence intervals is $R_{cpu}$ for $\beta$ = 8.
        
    \item 
        \textbf{Memory ($R_{mem}; \beta = 7$).} Regarding memory, Graph~(b) of Fig.~\ref{fig:lin_exp_inline}, it considers fixed mean values of  1.2 \ac{vcpu} cores and 600\,Mbps of \ac{lc}. 
        $R_{mem}$ with a different $\beta$ parameter value settings allocates converges to approximately 1123\,MB after approximately 1200 episodes. 
        The best reward function based on the smallest confidence intervals is $R_{mem}$ for $\beta$ = 7.

    \item 
        \textbf{Link Capacity ($R_{lc}; \beta = 7$).} We study \ac{lc} for 1.2 \ac{vcpu} cores and 1300 MB of memory. 
        As portrayed in graph~\ref{fig:lin_exp_inline}(c), the best reward option is $R_{lc}$ with  steepness coefficient
        $\beta = 7$, which yields an LC approximately equal to 475\,Mbps whereas converging to this optimal final
        state faster than alternatives after episode 1250. 
        
\end{itemize}

\subsubsection{\snort (Passive mode)}

\begin{figure*}[h]
    \centering 
    \begin{subfigure}[b]{0.45\columnwidth}
        \includegraphics[width=\textwidth]{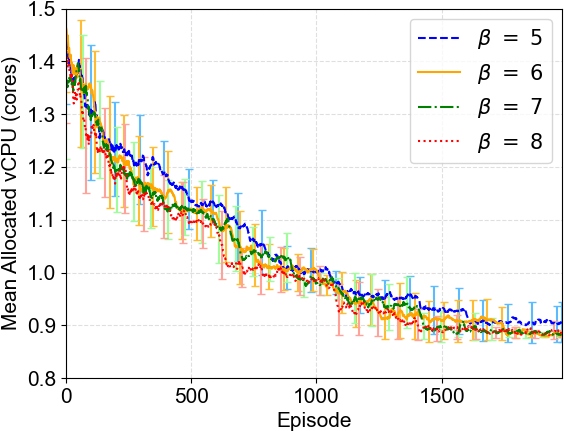}
        \caption{}
        \label{fig:lin_exp_passive_a}
    \end{subfigure}
    \begin{subfigure}[b]{0.45\columnwidth}
        \includegraphics[width=\textwidth]{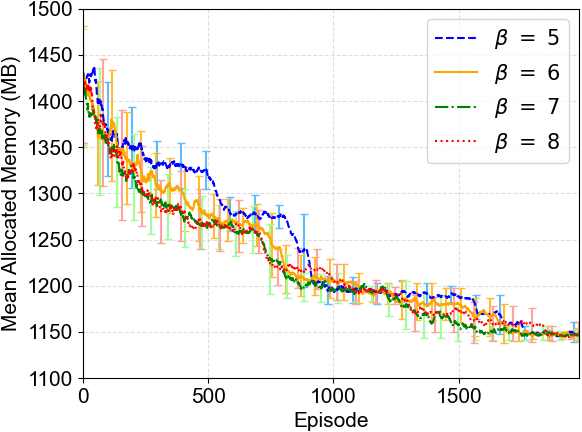}
        \caption{}
        \label{fig:lin_exp_passive_b}
    \end{subfigure}
    \begin{subfigure}[b]{0.45\columnwidth}
        \includegraphics[width=\textwidth]{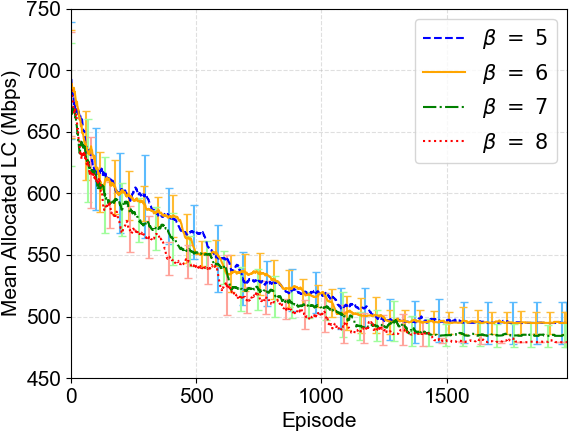}
        \caption{}
        \label{fig:lin_exp_passive_c}
    \end{subfigure}
    \caption{Configuring reward function parameters for \snort with Passive mode.}
    \label{fig:lin_exp_passive}
\end{figure*}

\begin{itemize}
    \item 
        \textbf{vCPU ($R_{cpu}; \beta = 8$).} 
         Fig.~\ref{fig:lin_exp_passive}a shows the results of an experiment in which the model is trained to adapt the number of \ac{vcpu} cores from 0.6 to 1.8 whereas keeping the values of memory and \ac{lc} fixed at 1300 MB and 600 Mbps, respectively. The $R_{cpu}$ decreases from 1.44 to about 0.88 at episode 1500 
         in all the alternatives of the steepness coefficient values ($\beta$). However, the best reward option is $R_{cpu}$, $\beta = 8$ as it converges to the final state faster than alternatives after approximately 1400 episodes.
         
    \item 
        \textbf{Memory ($R_{mem}; \beta = 7$).} In Fig.~\ref{fig:lin_exp_passive}(b),  
        $R_{mem}$ graph scales down 
        from 1425 MB to 1147 MB assuming fixed mean values of 1.2 \ac{vcpu} cores and 600 Mbps of \ac{lc} after approximately 1600 episodes. The most effective reward function for minimising memory is when $\beta = 7$ based on the smallest convergence time.  
        
    \item 
        \textbf{Link Capacity ($R_{lc}; \beta = 8$).} Fig.~\ref{fig:lin_exp_passive}(c) shows the mean allocated \ac{lc} for 1.2 \ac{vcpu} cores and 1300\,MB of memory. However, the best reward function based on minimum \ac{lc} is for $\beta = 9$ where 
        \ac{lc} can be reduced from 689 Mbps to approximately 480 Mbps at episode 1525.
\end{itemize}

\subsubsection{Virtual FireWall}

\begin{figure*}[h]
    \centering 
    \begin{subfigure}[b]{0.45\columnwidth}
        \includegraphics[width=\textwidth]{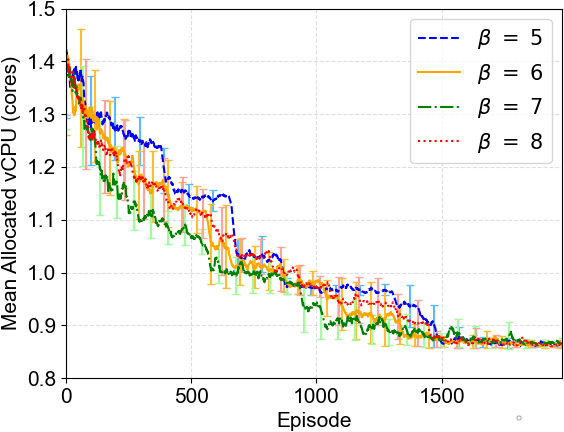}
        \caption{}
        \label{fig:lin_exp_firewall_a}
    \end{subfigure}
    \begin{subfigure}[b]{0.45\columnwidth}
        \includegraphics[width=\textwidth]{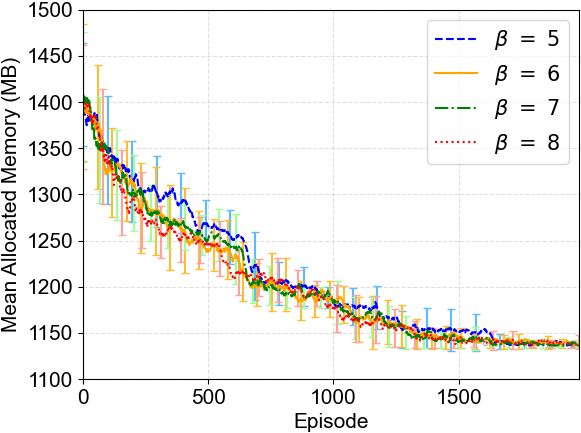}
        \caption{}
        \label{fig:lin_exp_firewall_b}
    \end{subfigure}
    \begin{subfigure}[b]{0.45\columnwidth}
        \includegraphics[width=\textwidth]{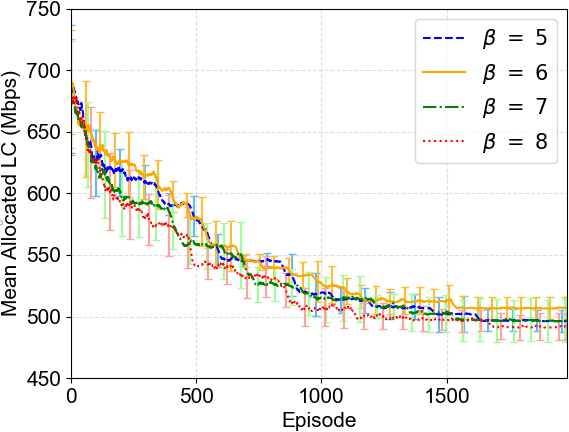}
        \caption{}
        \label{fig:lin_exp_firewall_c}
    \end{subfigure}
    \caption{Configuring reward function parameters for \ac{vfw}.}
    \label{fig:lin_exp_firewall}
\end{figure*}

\begin{itemize}
    \item 
        \textbf{vCPU ($R_{cpu}; \beta = 7$).} 
    Fig.~\ref{fig:lin_exp_firewall}(a) demonstrates an experiment where the model was trained to vary \ac{vcpu} values from 0.6 to 1.8 cores while maintaining memory and \ac{lc} at 1300 MB and 600 Mbps, respectively.
    Using $R_{cpu}$, it can be seen that the \ac{vcpu} cores get reduced from 1.40 to approximately 0.87 at episode 1450. 
    However, we find the minimum \ac{vcpu} cores faster using $\beta = 7$.
    \item 
        \textbf{Memory ($R_{mem}; \beta = 7$).} 
        For $R_{mem}$ in Fig.~\ref{fig:lin_exp_firewall}(b), the system using the $R_{mem}$ reward function, tries to make adjustments to decrease the memory from 1430 to 1140 MB when the vcpu cores are 1.2 and \ac{lc} is 600 Mbps. 
        However, the best reward function based on the minimum memory is $R_{mem}$ for $\beta = 7$ at episode 1600.        
    \item 
        \textbf{Link Capacity ($R_{lc}; \beta = 9$).} 
        As shown in Fig.~\ref{fig:lin_exp_firewall}(c), we use $R_{lc}$ to find the minimum \ac{lc} where \ac{vcpu} cores  are 1.2 cores and memory is 1300 MB. However, $\beta = 9$ can significantly reduce \ac{lc} from 686 Mbps to 482 Mbps. 
\end{itemize}

\vspace{-0.25cm}
\section{Performance study} 
\label{sec:eval}

We conduct a comprehensive search in Sec.~\ref{sec:eval:exaustive} 
to discover an ``Oracle" model of optimal profiles in a simulation environment. These optimal solutions set the ultimate performance targets for our \ac{rl} model. 
Moreover, a practical assessment of our approach requires a comparison against intelligent models thus we train \ac{sl} models  
and compare their performance against online learning over a \textit{dynamic environment} with growing dataset size, {\Revised so as to draw adaptability conclusions}. 

\vspace{-0.125cm}\vspace{-0.25cm}
\subsection{Oracle resource allocation (exhaustive search study)} 
\label{sec:eval:exaustive}

The results presented in Fig.~\ref{fig:corr_inline},\,\ref{fig:corr_passive}~and\,\ref{fig:corr_vFW} correspond to each \vnf type, namely \snort for Inline mode, \snort for Passive mode and \ac{vfw}, respectively. 
Each figure contains 5 graphs that portray performance after an \textit{exhaustive} exploration of resource allocation combinations towards identifying an \textit{optimal trade-off} combining a minimum of resources for optimal performance in terms of \ac{or}. 
All performance measurements are based on the mean values of at least \textit{30 recorded instances} from a dataset attained during the offline profiling stage, alongside corresponding  \textit{95\%} confidence intervals. 
%
    {\Revised Graphs (a) and (b) in Figures~\ref{fig:corr_inline},\,\ref{fig:corr_passive}~and \,\ref{fig:corr_vFW}} show the mean \ac{or} against the number of allocated \ac{vcpu} cores and \ac{lc}, respectively, for different allocated memory levels mapped to each curve in the graphs {\Revised per each \ac{vnf} type}. Their purpose is to \textit{pinpoint a minimum} of resource allocation on the x-axis for which the \ac{or} on the y-axis converges to a maximum mean value. Specifically, Graph (a) in each figure above illustrates the impact of \ac{vcpu} cores on \ac{or} with a fixed \ac{lc} of 600 Mbps, while Graph~(b) shows the impact of \ac{lc} with a fixed allocation of 1.2 cores for vCPUs. Fixing these values serves to focus on the direct relationship between pairs of values. Note that fixed values are carefully selected to accommodate Optimum \acp{or} after preliminary test runs.
    Graphs (c) and (d) plot mean \ac{or} (orange curves) compared to consistently increasing \ac{lc} levels (blue curves). The y-axes show bit-rates against increasing \ac{lc} levels grouped by increasing \ac{vcpu} cores or memory for (c) and (d), respectively. If these two curves \textit{identify}, then the \ac{lc} is \textit{best-utilised}, with the best resource combinations achieved at optimal (i.e., maximised) \ac{or} levels. As with (a) and (b), we keep memory fixed at 1300\,MB for (c) and \ac{vcpu} at 1.2 cores for (d). 
    Last, Graph~(e) in each figure shows on the x-axis increasing memory levels grouped by incrementally increasing \ac{vcpu} cores: 0.6, 0.8, \dots, 1.8, given fixed 600\,Mbps.

\subsubsection{\snort (Inline mode)} 

\begin{figure*}[h]
\centering
\begin{subfigure}{0.45\textwidth}
  \includegraphics[width=.95\linewidth]{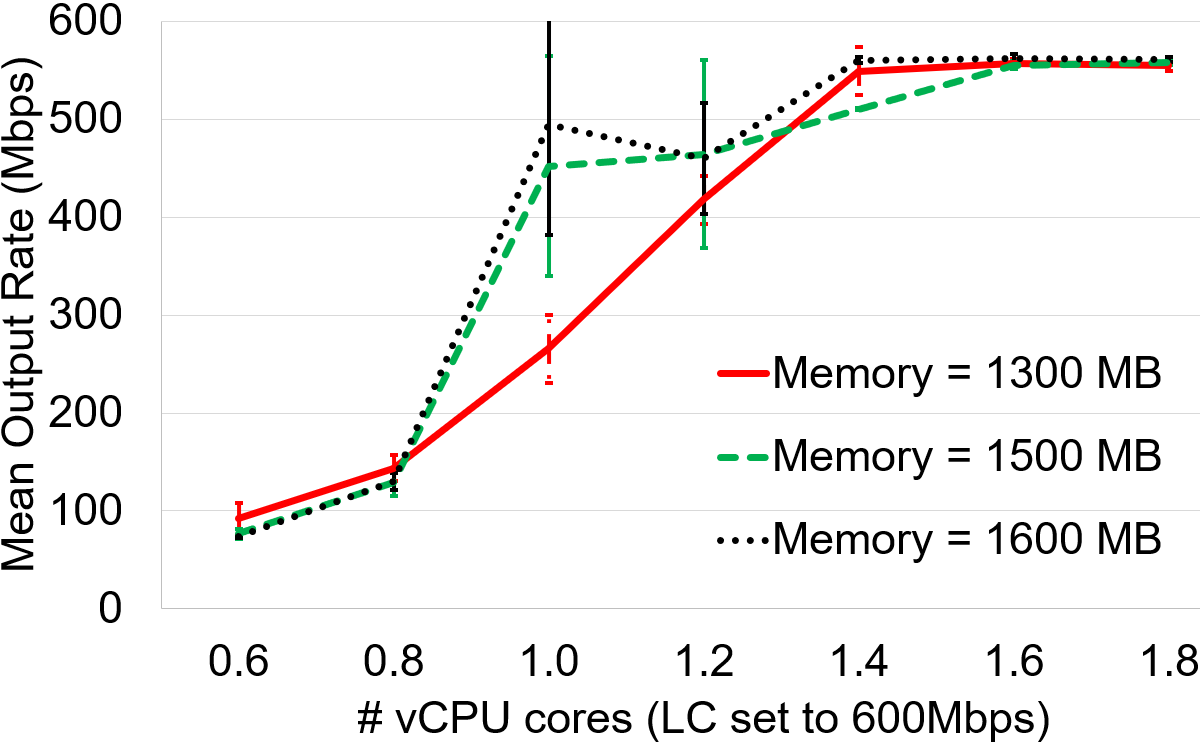}
  \caption{}
\end{subfigure}%
\begin{subfigure}{0.45\textwidth}
  \includegraphics[width=.955\linewidth]{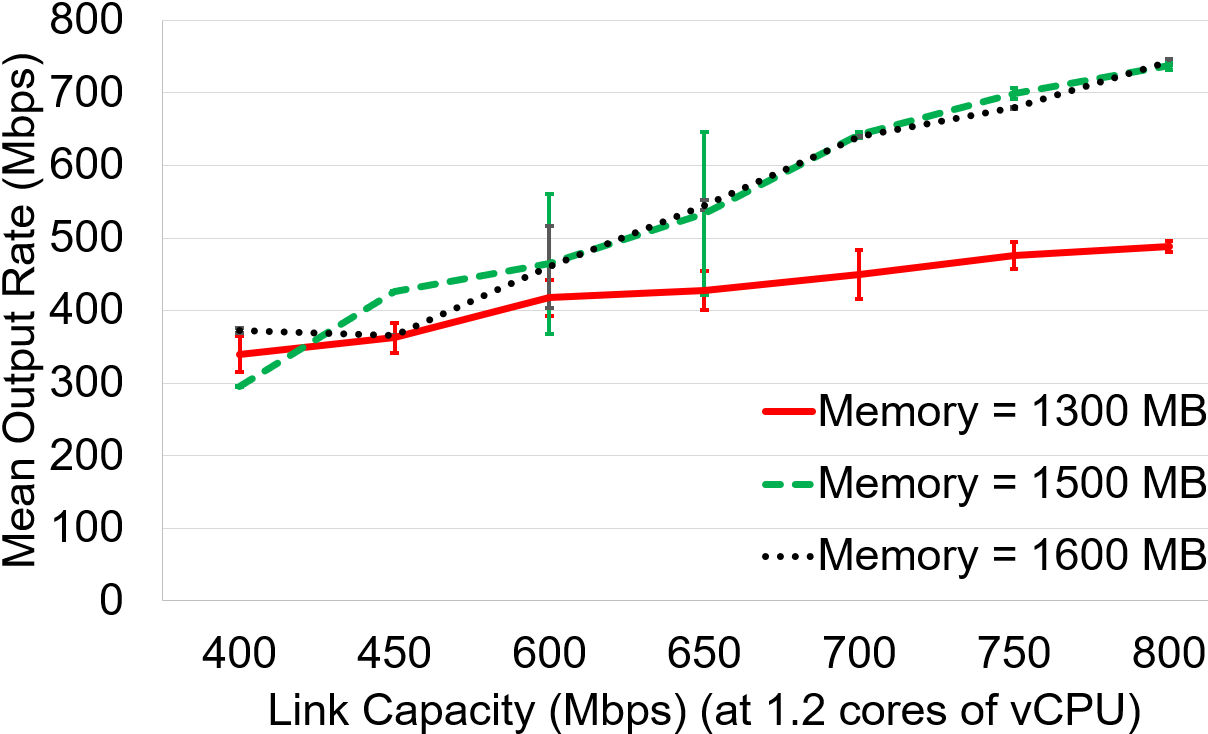}
  \caption{}
\end{subfigure}\par\medskip
\begin{subfigure}{0.45\textwidth}
  \includegraphics[width=0.95\linewidth]{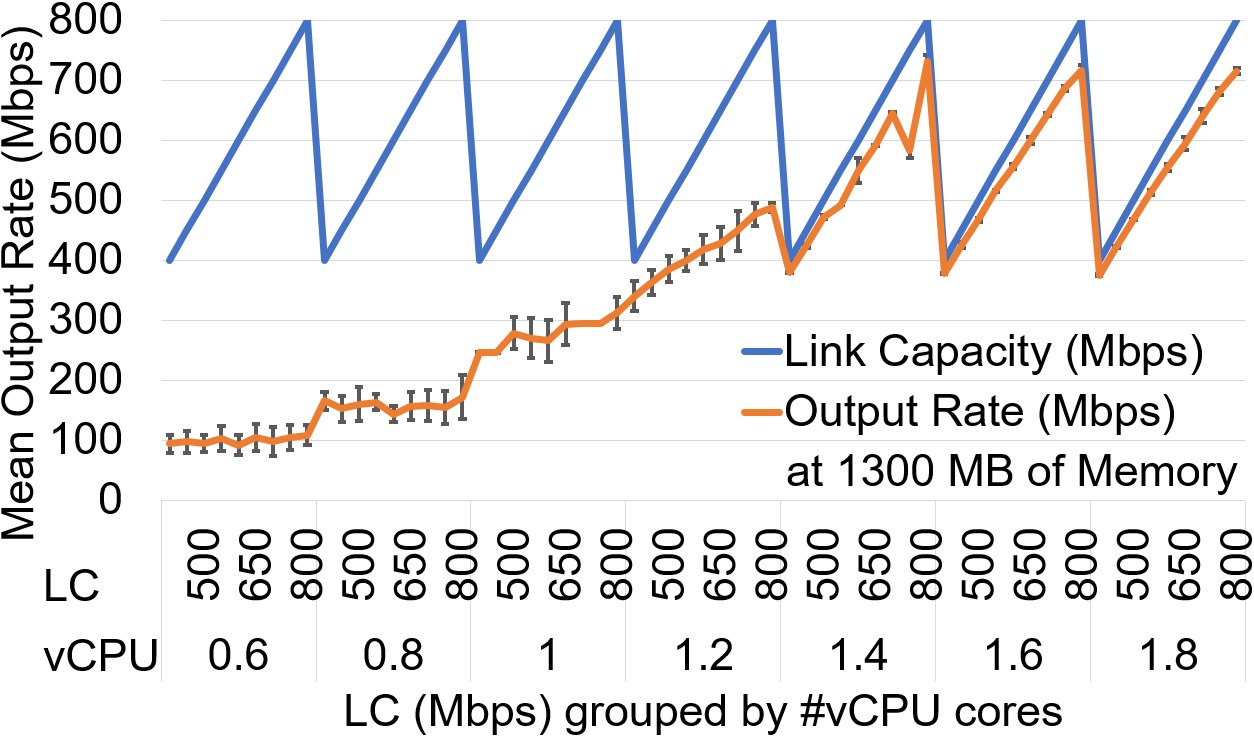}
  \caption{}
\end{subfigure}%
\begin{subfigure}{0.45\textwidth}
  \includegraphics[width=.95\linewidth]{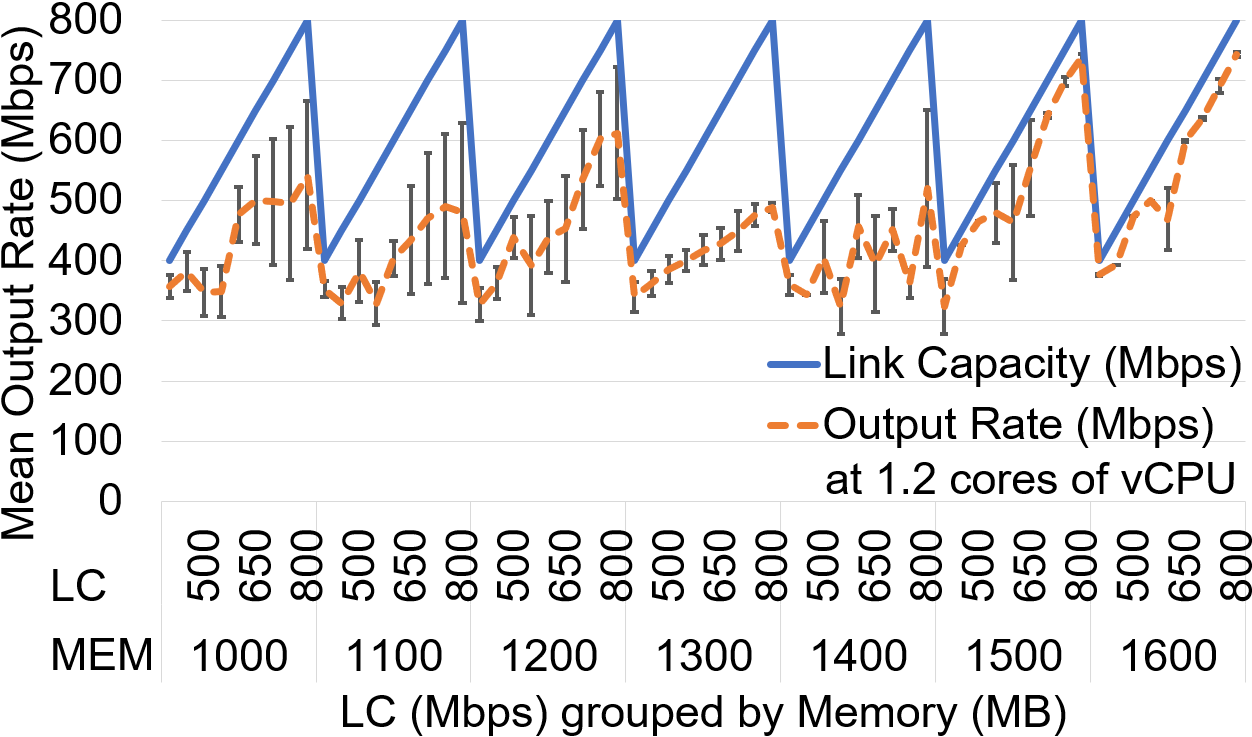}
  \caption{}
\end{subfigure}%
\\
\begin{subfigure}{0.45\textwidth}
  \includegraphics[width=0.95\linewidth]{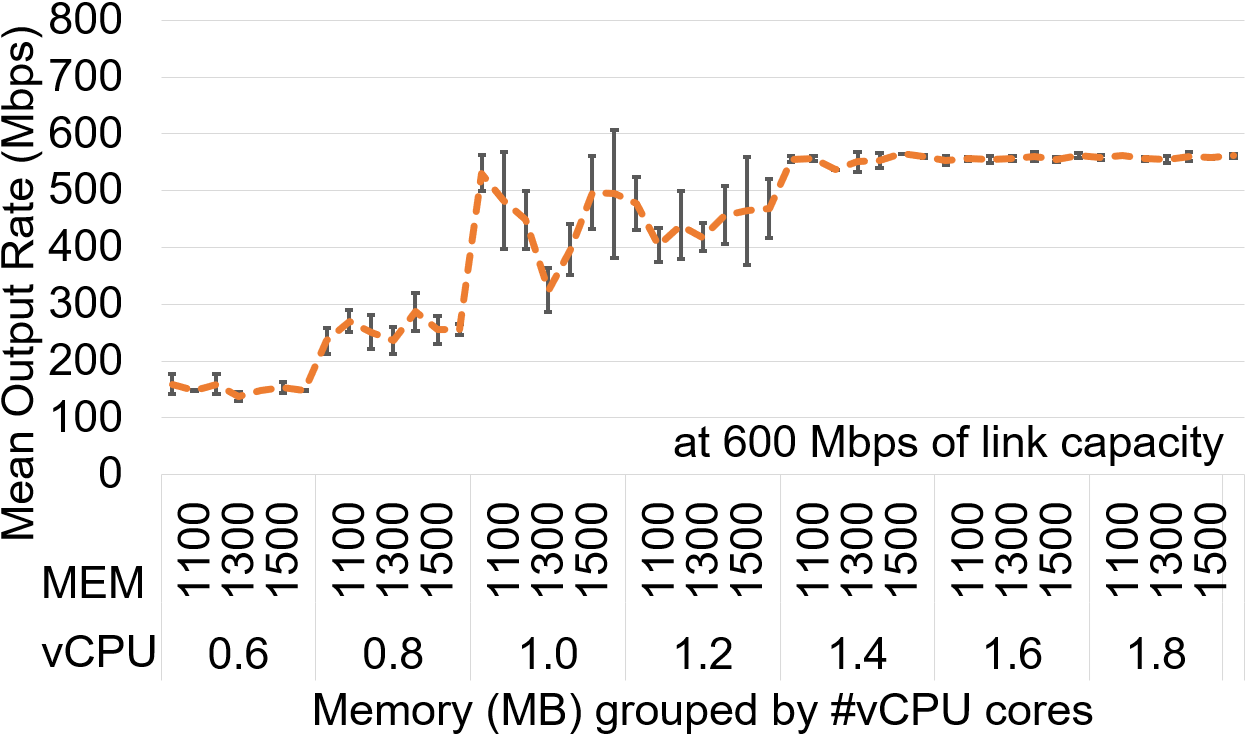}
  \caption{}
\end{subfigure}
\caption{Oracle resource allocation using \snort (Inline mode).}
\label{fig:corr_inline}
\end{figure*}

The graphs of Fig.\,\ref{fig:corr_inline}(a) and Fig.\,\ref{fig:corr_inline}(b) show three curves corresponding to 1300\,MB, 1500\,MB and 1600\,MB memory levels. 
   We observe that the \ac{or} grows with the number of \ac{vcpu} cores in the range of \textit{0.6 – 1.4} cores regardless of allocated memory in Fig.~\ref{fig:corr_inline}(a), excluding the case of 1.0 - 1.2 \ac{vcpu} for the 1500\,MB and 1600\,MB memory curves due to outliers as denoted by confidence intervals. 
   The \ac{or} also increases with \ac{lc} in the Graph~\ref{fig:corr_inline}(b) 
   for all memory curves. By comparing the different memory curves, increased memory results in a higher \ac{or}. 
    Based on the above, we can conclude that 
            the total allocation of \textit{all resource types} collectively affects the \ac{or}. 
            Also, \emph{\ac{or} converges} to a maximum of \textasciitilde 550\,Mbps  after $vCPU=1.4$.
        Another important conclusion from Graph~(b) (also backed by conclusions below after Graph~(d)) is that the \ac{or} for a memory of less than 1500\,MB ceases to increase and is, thus, \emph{sub-optimal}. At the same time, an increased memory allocation at 1600\,MB does not increase \ac{or} further. 
    Regarding Graphs~(c) and (d) of Fig.~\ref{fig:corr_inline}  
         the optimal utilisation of \ac{lc} can be achieved with \emph{minimum \ac{vcpu} 1.4}, as \ac{or} for increasing \lc in Graph (c) slowly converges and \textit{finally identifies with \lc} at a minimum (i.e., optimal) allocation of \ac{vcpu} 1.4. Note that this is consistent with the observation from Graph \ref{fig:corr_inline}(a) (see above).
        The \textit{best \ac{lc} utilisation can be achieved with a minimum of memory} (1500\,MB), as \ac{or} in Graph (d) for increasing \lc identifies with \lc for a minimum (i.e., optimal) memory level of 1500. For completeness, we note that lower memory levels like for 1100\,MB show a linear (but not identifying) trend between \ac{or} and \ac{lc} curves, yet with large confidence intervals.
%
    Last, Graph~(e) of Fig.~\ref{fig:corr_inline} leads to the conclusion that
            \ac{or} (orange curve) generally \textit{increases with \ac{vcpu}} until before $vCPU=1.4$ irrespective of some large confidence interval values, \textit{and then converges} for \ac{vcpu}$\geq$1.4.
            This is consistent with the observation from Graph~\ref{fig:corr_inline}(a), and with the conclusion from Graph~\ref{fig:corr_inline}(c).
\subsubsection{\snort (Passive mode)}

Likewise to \snort for Inline mode, the conclusions for each graph of Fig.~\ref{fig:corr_passive} are as follows.
\begin{figure*}[h]
\centering
\begin{subfigure}{0.45\textwidth}
  \includegraphics[width=.925\linewidth]{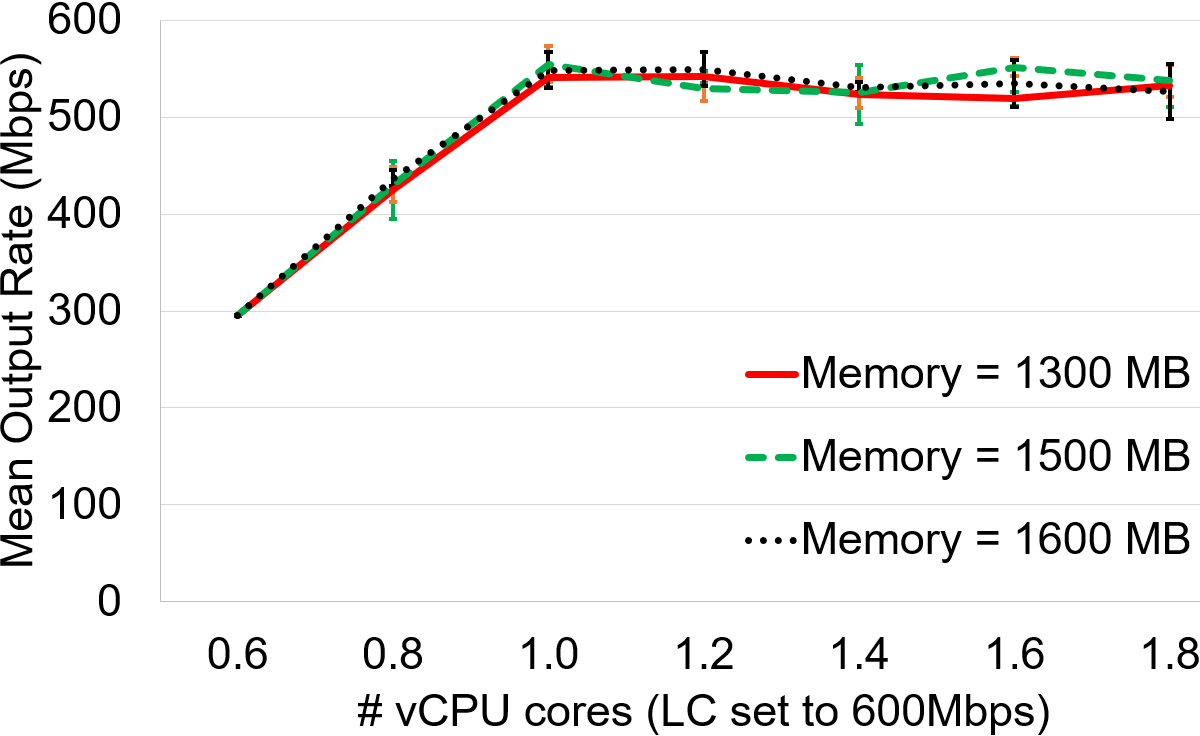}
  \caption{}
\end{subfigure}%
\begin{subfigure}{0.45\textwidth}
  \includegraphics[width=.9255\linewidth]{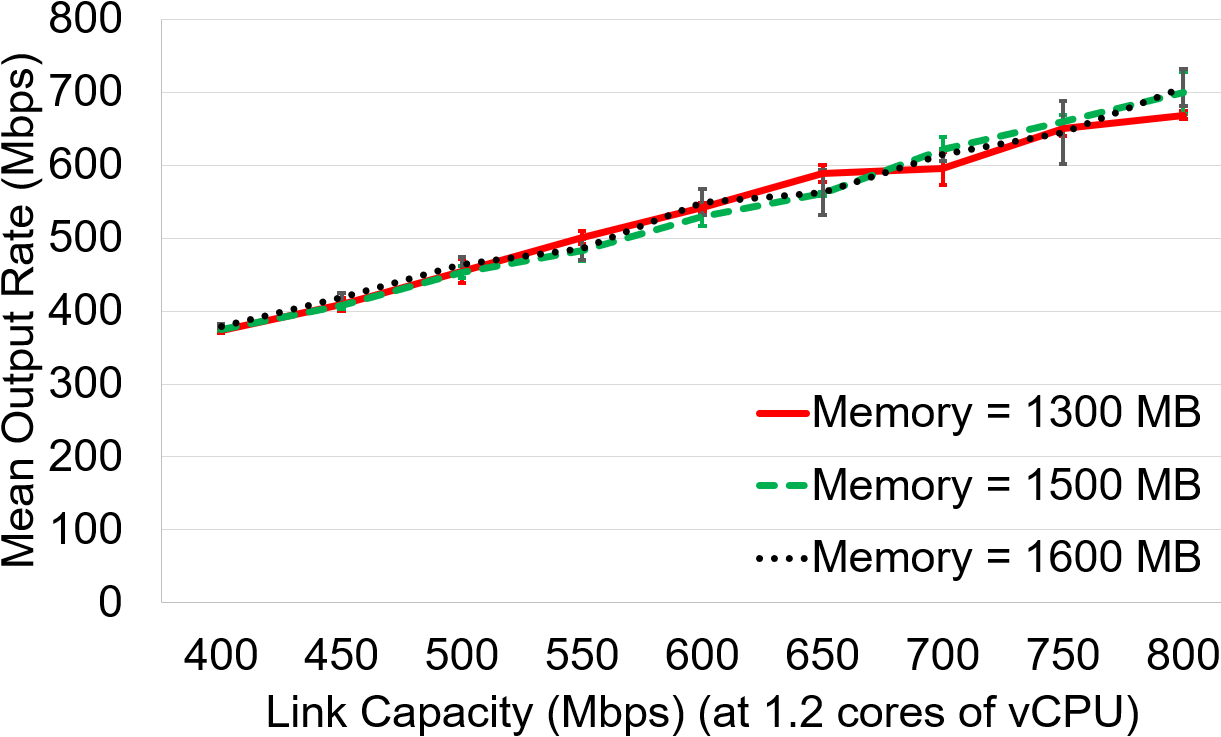}
  \caption{}
\end{subfigure}\par\medskip
\begin{subfigure}{0.45\textwidth}
  \includegraphics[width=0.925\linewidth]{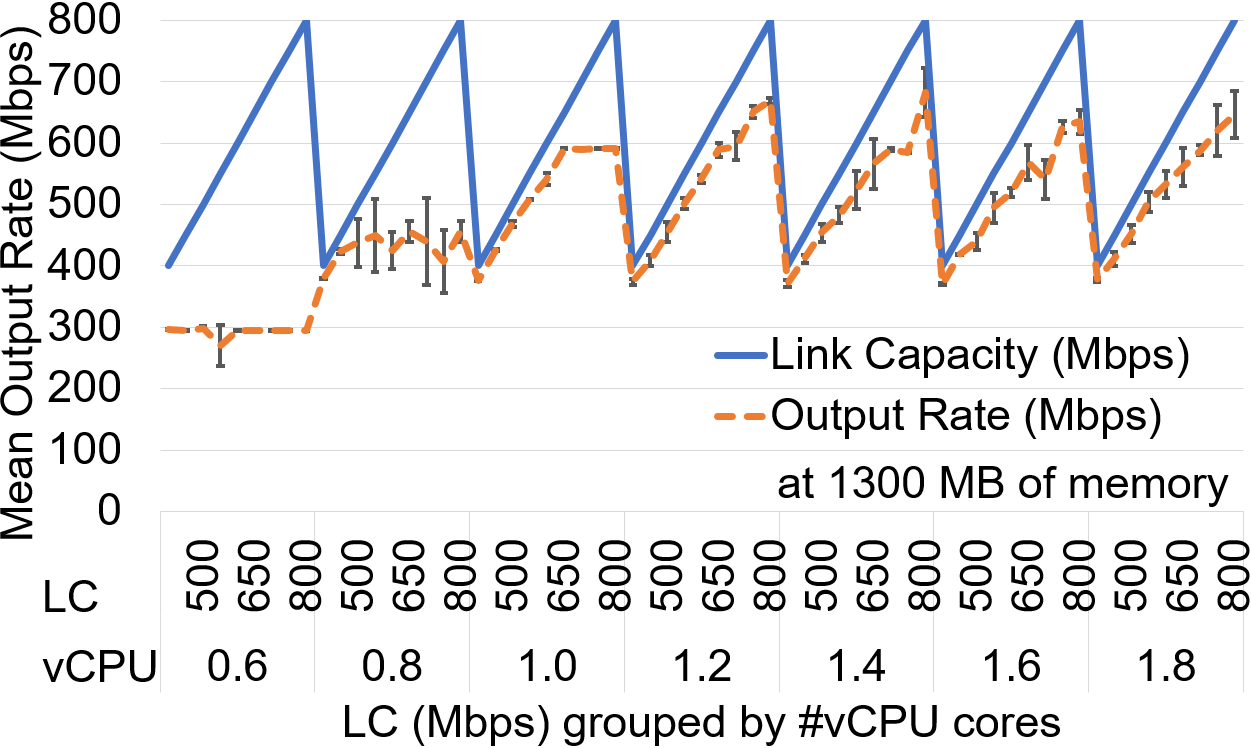}
  \caption{}
\end{subfigure}%
\begin{subfigure}{0.45\textwidth}
  \includegraphics[width=.925\linewidth]{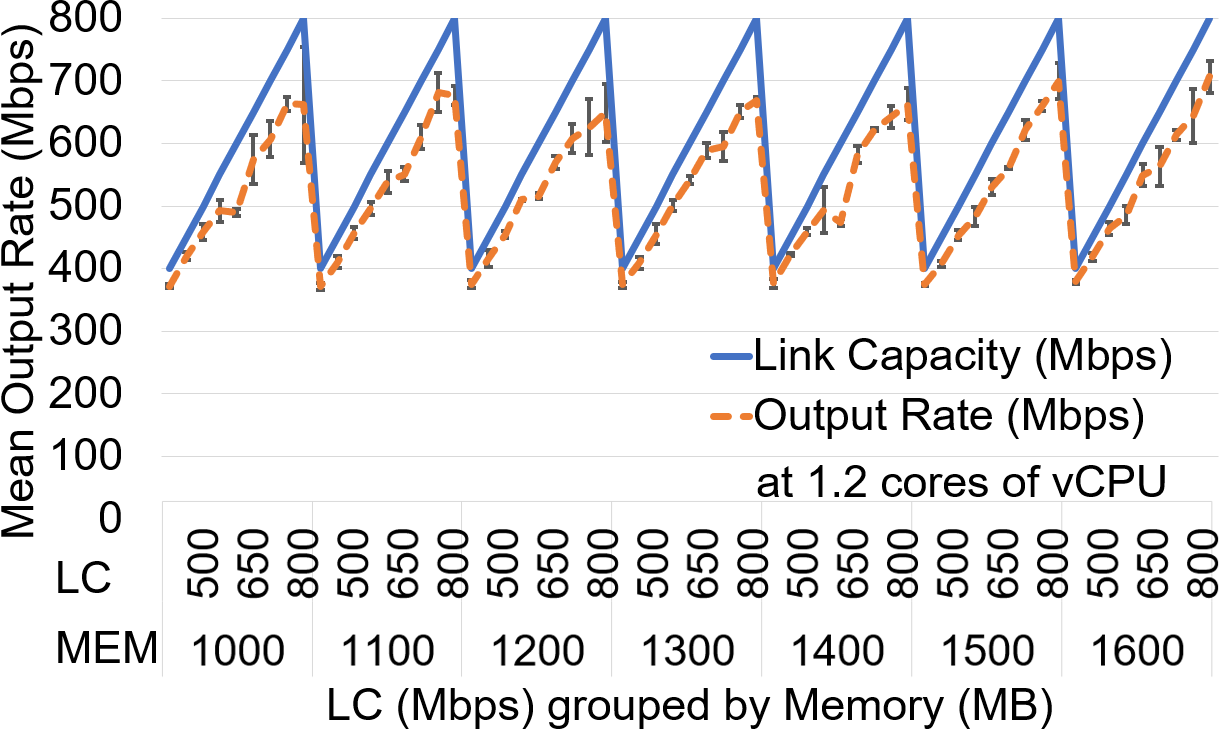}
  \caption{}
\end{subfigure}
\\
\begin{subfigure}{0.45\textwidth}
  \includegraphics[width=0.925\linewidth]{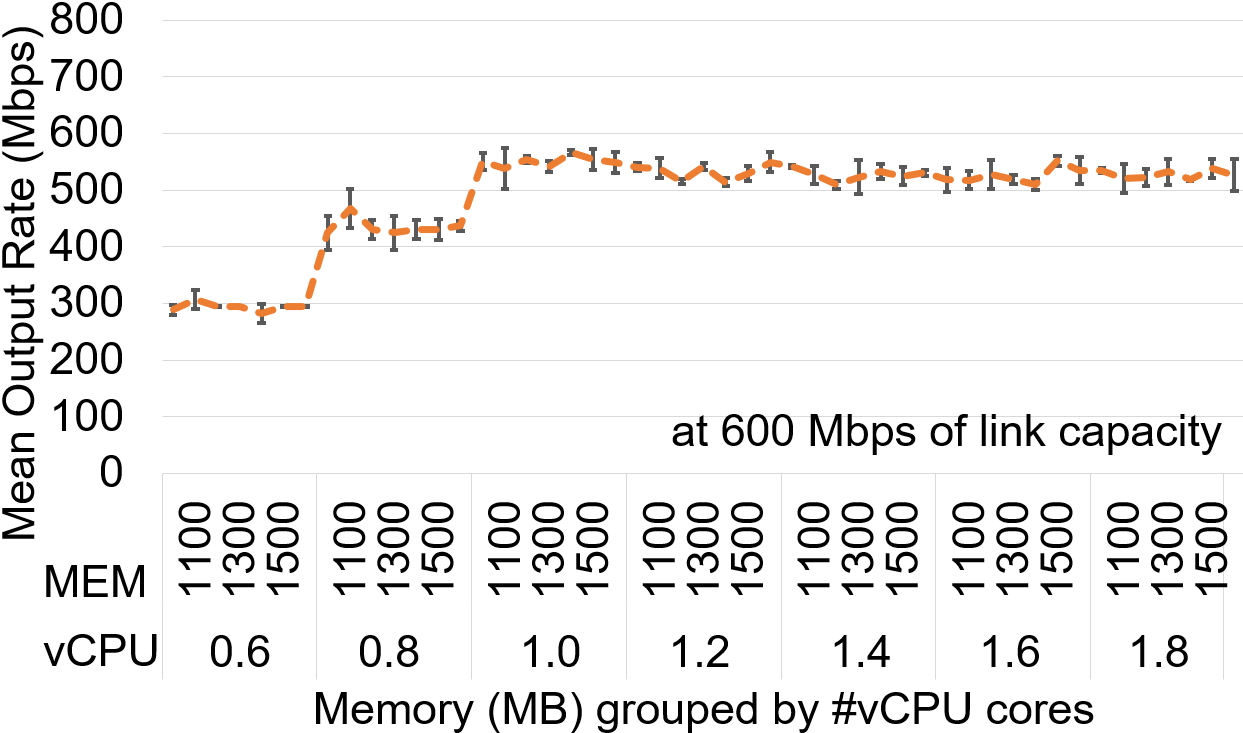}
  \caption{}
\end{subfigure}
\caption{Oracle resource allocation using \snort (Passive mode).}
\label{fig:corr_passive}
\end{figure*}
Regarding Graph~(a), increasing \ac{vcpu} cores causes a higher \ac{or}. However, the \ac{or} converges and remains at around 525 Mbps in the range of 1.0 – 1.8  \ac{vcpu} cores. This holds for all memory level curves, for which \lc results strongly identify.
Graph (b), on the other hand, shows that the \ac{or} increases in an almost linear function with \ac{lc} at 1.2 \ac{vcpu} cores at all memory sizes. In addition, Even though we added more memory across all \ac{vcpu} cores, as also shown in Graph~(e), the \ac{or} at each \ac{vcpu} core remained the same. Therefore we conclude that
    \textit{memory does not impact  \ac{or}}. This is due to this \ac{vnf} type's different nature compared to Snort (Inline mode), with the latter needing memory resources to inspect packets before forwarding them.
 %
The \ac{or} in Graph~(c) is similar to \ac{lc} in the range of 1.0-1.8 of \ac{vcpu} cores, while in Graph~(d) \ac{or} changes along with \ac{lc} across the memory range. We conclude that the \ac{vcpu} cores and \ac{lc} affect \ac{or}, but memory does \textit{not}.

\subsubsection{virtual FireWall}

The graphs of  Fig.~\ref{fig:corr_vFW} for the case of \ac{vfw} are similar to the ones for \snort (Passive mode), where the \ac{or} depends on the \ac{lc} across the \ac{vcpu} core and memory range. 
Nevertheless, for \ac{vcpu} equal to 0.6 in Graphs~(a) and  (c), the \ac{or} also seems to depend on the \ac{vcpu} core. Moreover, Graphs (a), (b), and (e) show that the output value does not change even when the memory is increased. And last, the \ac{lc} obviously affects the \ac{or}, as shown in Graphs (b) and (d).
\begin{figure*}[h]
\centering
    \begin{subfigure}{0.45\textwidth}
      \includegraphics[width=.95\linewidth]{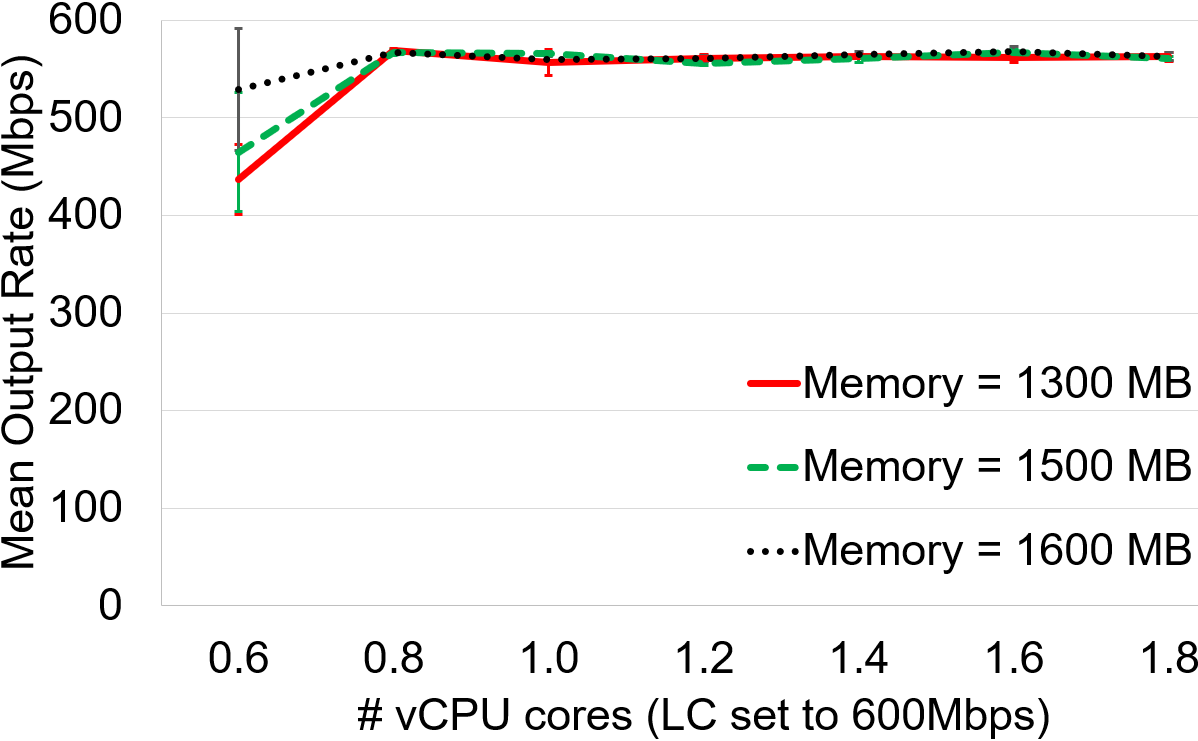}
      \caption{}
    \end{subfigure}%
    \begin{subfigure}{0.45\textwidth}
      \includegraphics[width=.955\linewidth]{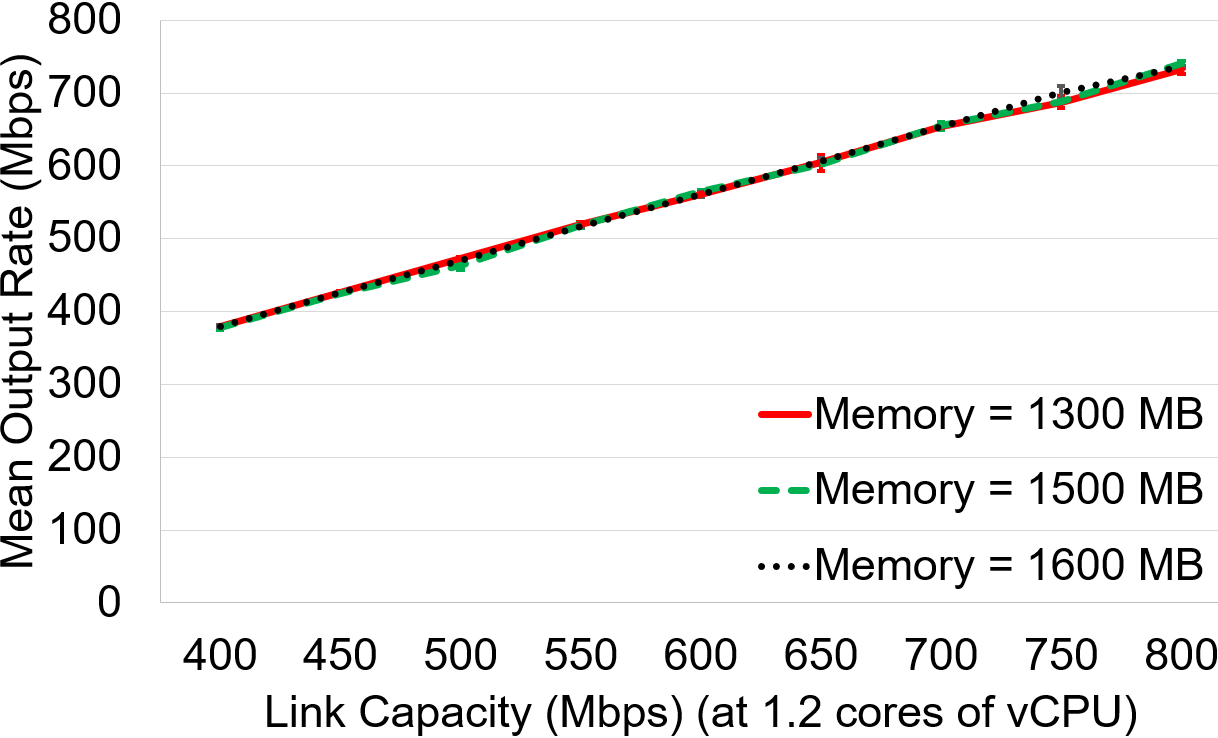}
      \caption{}
    \end{subfigure}\par\medskip
    \begin{subfigure}{0.45\textwidth}
      \includegraphics[width=0.95\linewidth]{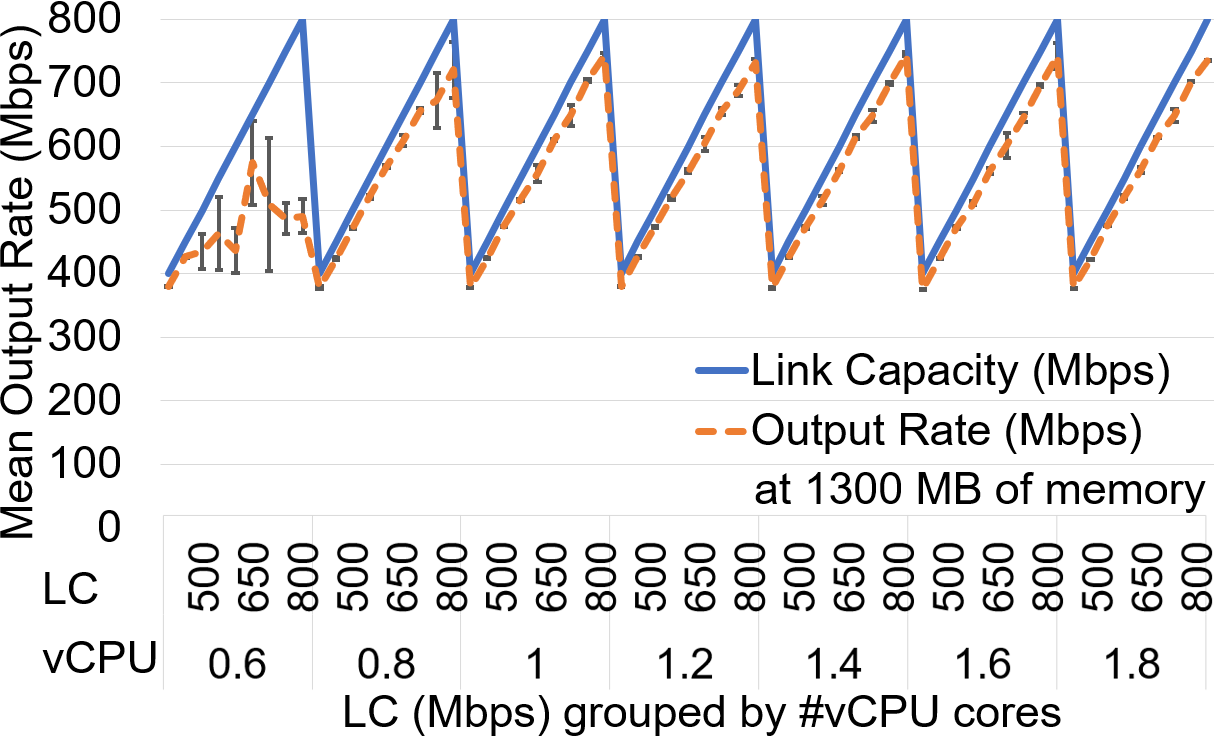}
      \caption{}
    \end{subfigure}%
    \begin{subfigure}{0.45\textwidth}
      \includegraphics[width=.98\linewidth]{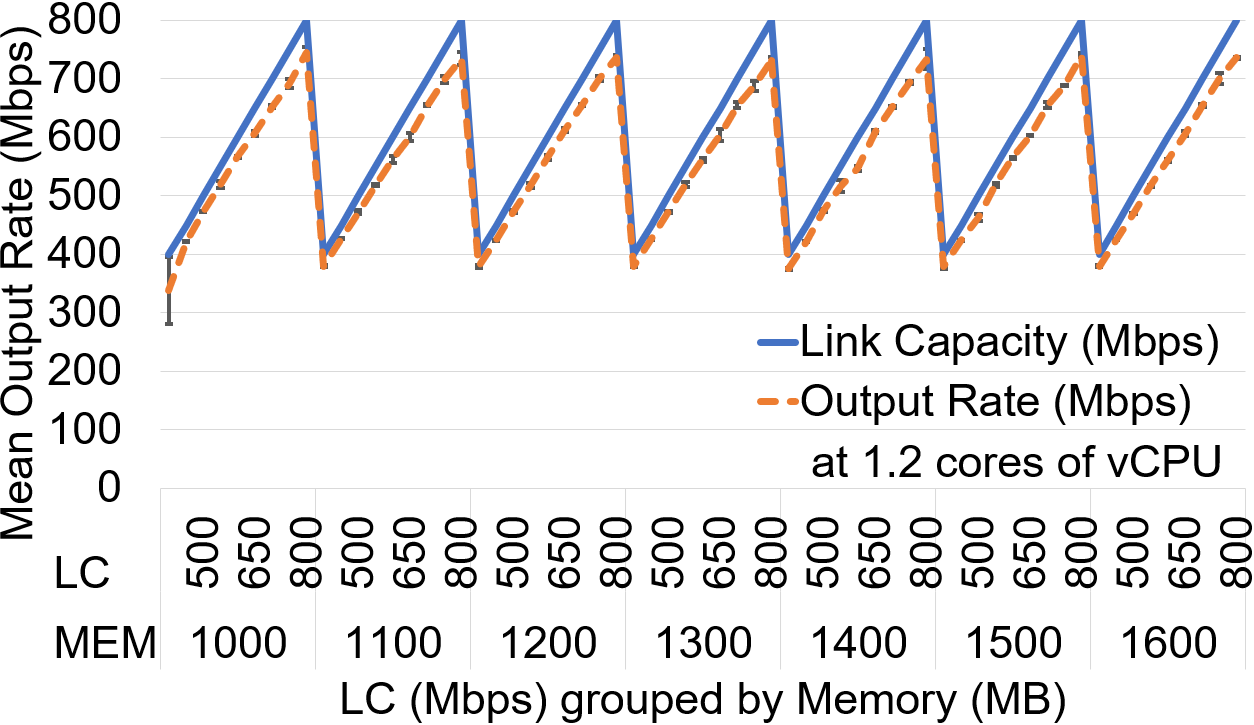}
      \caption{}
    \end{subfigure}%
    \\
    \begin{subfigure}{0.45\textwidth}
      \includegraphics[width=0.96\linewidth]{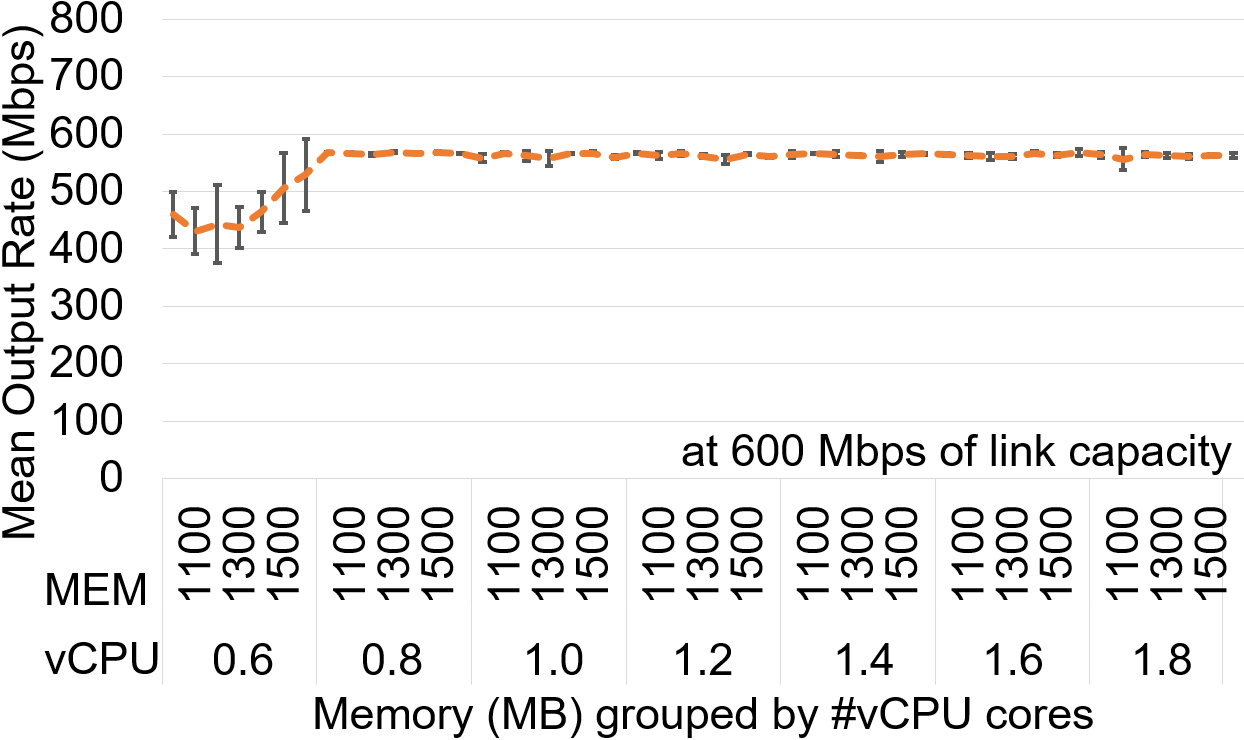}
      \caption{}
    \end{subfigure}
\caption{Oracle resource allocation using \ac{vfw}.}
\label{fig:corr_vFW}
\end{figure*}

\subsection{Online learning profiling performance} 
\label{sec:eval:results}

    We assess iOn-Profiler's Q-Learning adaptation of Algorithm~\ref{alg:scale_multiobj} in a dynamic environment where the dataset size \textit{grows} at run time. We compare the predicted resources 
    to those obtained in a static environment with a static dataset for each training episode, as shown in Fig.~\ref{fig:predict_inline} and Fig.~\ref{fig:predict_passive}. 
    We calculate the percentage error in resource allocation compared to the optimal allocation, along with 95\% confidence intervals for each resource type and the reported results refer to a scenario of\footnote{
        We elaborate on this in Sec.~\ref{sec:combine_weight}, where scenario w(\sfrac{1}{3}, \sfrac{1}{3}, \sfrac{1}{3}) is only one out of 13 for each of the 3 \ac{vnf} types outlined in Tables~\ref{table:output_steady_state_vfw}, ~\ref{table:output_steady_state_inline} and ~\ref{table:output_steady_state_passive}.    
    } 
    \textit{equal resource} weights w(\sfrac{1}{3}, \sfrac{1}{3}, \sfrac{1}{3}) and tuned parameters $\beta$ for each resource in a static environment. 

\subsubsection{Setup and training of SL benchmarks}
\label{sec:eval:slbenchmarks}

{\Revised 
  For the \ac{mlp} and \ac{rf} benchmarks, we forecast the resources required periodically at landmark episodes (where \ac{rl} performance is recorded and depicted in Figures~\ref{fig:predict_inline}-\ref{fig:predict_vFW}) by training the models on the available dataset collected up to that episode.
  %
   Then, we split the dataset into a 90:10 ratio for the training and test sets, and normalised it using min-max feature scaling. %
   This approach allows us to conduct a fair comparison between RL and the benchmarks over the same training data.}
   
   The input variables for the \ac{sl} predictions include \ac{vcpu} and memory utilisation, latency, and Optimum \ac{or}, while output variables include the number of \ac{vcpu} cores,  memory, and \ac{lc}. The number of trees in the \ac{rf} is set to 500, 500, and 800 for Snort (Inline mode), Snort (Passive mode) and \ac{vfw}, respectively. The \ac{mlp} parameters are described in Tab.\,\ref{table:paramMLP}.
{
    \begin{table}[h]
    \centering
    \caption{Parameters of the MLP Model}
    \label{table:paramMLP}
    \begin{tabular}{lccc}
    \toprule
    \multicolumn{1}{c}{Parameter}                          & \begin{tabular}[c]{@{}c@{}}\snort\\ (Inline)\end{tabular} & \begin{tabular}[c]{@{}c@{}}\snort\\ (Passive)\end{tabular} & vFW                       \\ 
    \midrule
    \multicolumn{1}{l}{Number of neurons in Input Layer} & \multicolumn{1}{c}{4}                                   & \multicolumn{1}{c}{4}                                    & \multicolumn{1}{c}{4}    \\
    Number of neurons in Output Layer                      & 3                                                        & 3                                                         & 3                         \\
    Number of neurons in 1st Hidden Layer                  & 512                                                      & 256                                                       & 128                       \\
    Number of neurons in 2nd Hidden Layer                  & 256                                                      & 128                                                       & 128                       \\
    Number of neurons in 3rd Hidden Layer                  & 256                                                      & 128                                                       & 128                       \\
    Activation Function in Hidden Layer                    & selu                                                     & selu                                                      & selu                      \\
    Activation Function in the Output Layer                & sigmoid                                                  & sigmoid                                                   & sigmoid                   \\
    Epoch                                                  & 500                                                      & 500                                                       & 500                       \\
    Batch size                                             & 16                                                       & 16                                                        & 16                        \\
    optimiser                                              & Adam                                                     & Adam                                                      & Adam                      \\
    \multicolumn{1}{l}{Learning rate}                    & \multicolumn{1}{c}{1e-4}                                & \multicolumn{1}{c}{1e-4}                                 & \multicolumn{1}{c}{1e-4} \\ 
    \bottomrule
    \end{tabular}
    \end{table}
}



\begin{figure*}[h]
    \centering 
    \begin{subfigure}{0.45\columnwidth}
        \includegraphics[width=\textwidth]{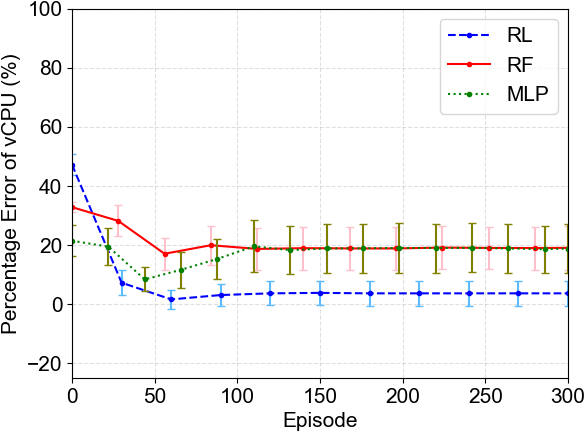}
        \caption{}
        \label{fig:error_inline_a}
    \end{subfigure}
    \begin{subfigure}{0.45\columnwidth}
        \includegraphics[width=\textwidth]{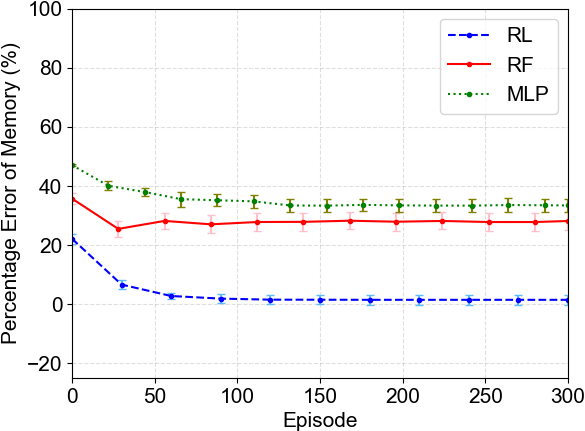}
        \caption{}
        \label{fig:error_inline_b}
    \end{subfigure}
    \begin{subfigure}{0.45\columnwidth}
        \includegraphics[width=\textwidth]{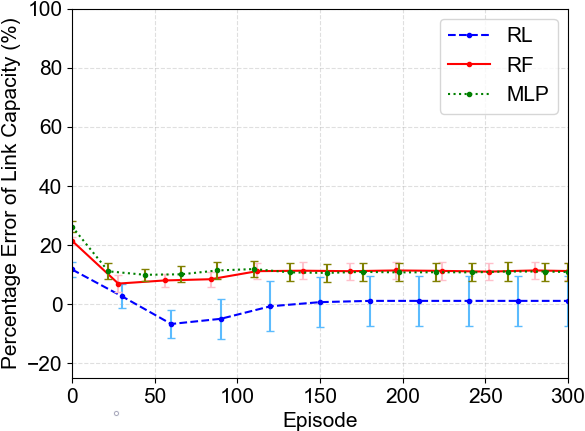}
        \caption{}
        \label{fig:error_inline_c}
    \end{subfigure}
    \caption{Percentage error of resource allocation predictions by \ac{mlp}, \ac{rf} and \ac{rl} regarding \snort (Inline mode).} 
    \label{fig:predict_inline}
\end{figure*}

\captionsetup{skip=0pt} 

\subsubsection{Snort (Inline mode)}
Graphs (a), (b), and (c) of Fig.~\ref{fig:predict_inline} show the resource allocation percentage error for \snort with Inline mode.
    According to (a) and (b), \rl has less \ac{vcpu} and memory percentage error than \ac{mlp} and \ac{rf}. As for Fig.~\ref{fig:predict_inline}(c), \ac{mlp} and \ac{rf} do not significantly reduce \ac{lc} whereas the \rl gives a lower percentage error. 
    We can infer from the data above that \rl can provide a lower percentage of prediction resource error than \ac{mlp} and \ac{rf}. The underlying reason is that \rl learns to reduce resource consumption from past events. In contrast, the resource allocation percentage error of \ac{mlp} and \ac{rf} are high because they use a static trained model that makes them unable to adapt to reduce resource consumption.

\begin{figure*}[h]
    \centering 
    \begin{subfigure}[hbt!]{0.45\columnwidth}
        \includegraphics[width=\textwidth]{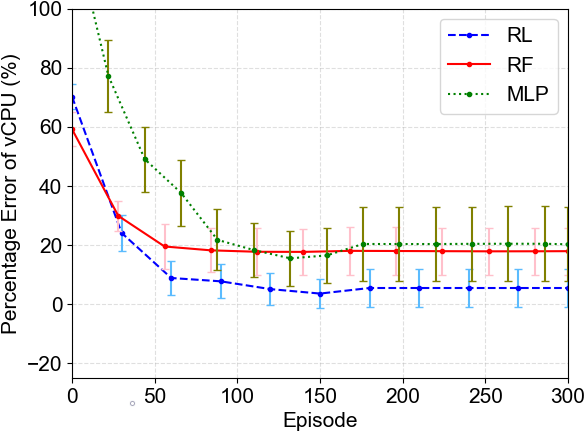}
        \caption{}
        \label{fig:error_passive_a}
    \end{subfigure}
    \begin{subfigure}[hbt!]{0.45\columnwidth}
        \includegraphics[width=\textwidth]{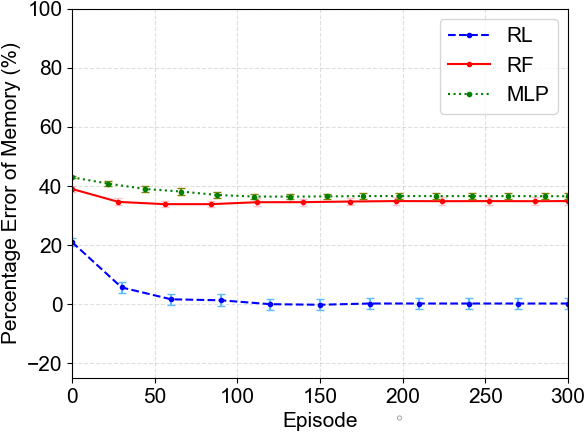}
        \caption{}
        \label{fig:error_passive_b}
    \end{subfigure}
    \begin{subfigure}[hbt!]{0.45\columnwidth}
        \includegraphics[width=\textwidth]{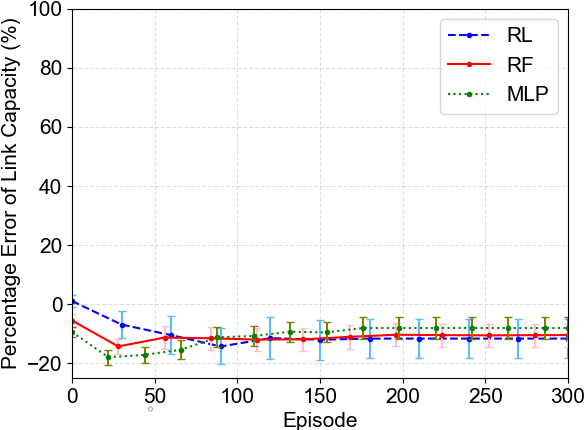}
        \caption{}
        \label{fig:error_passive_c}
    \end{subfigure}
    \caption{Percentage error of resource allocation predictions by \ac{mlp}, \ac{rf} and \ac{rl} regarding \snort (Passive mode).} 
    \label{fig:predict_passive}
\end{figure*}
\subsubsection{Snort (Passive mode)}
    According to Graph~(a) of Fig.~\ref{fig:predict_passive}, \rl produces a lower percentage error regarding \ac{vcpu} cores than \ac{mlp} and \ac{rf}. In terms of Memory, Graph (b) shows that \rl yields \emph{no} error contrary to \ac{mlp} and \ac{rf}. In Graph~(c), the \ac{lc} percentage error is negative and quite similar for \ac{rl}, \ac{mlp}, and \ac{rf}.
    Therefore, \rl is more accurate for \snort with Passive mode as it yields lower error percentages than \ac{mlp} and \ac{rf}, notably for \ac{vcpu} and memory.
    The analysis presented in Figure~\ref{fig:predict_vFW} focuses on the performance of the \ac{vfw} \ac{vnf}. In general, our online \ac{rl} model exhibits notably superior capabilities for predicting resource allocation compared to the benchmarks across all resources.
    When examining Graphs (a), (b), and (c) after 150-175 episodes\footnote{
        This range poses an approximate performance convergence milestone across all graphs and models. 
    }, 
    the \ac{rl} model demonstrates mean percentage errors of 9\%, 2\%, and 5\% for \ac{vcpu}
    cores, memory, and \ac{lc} respectively. In contrast, the \ac{mlp} and \ac{rf} models yield errors of 
    52\% and 18\% for \ac{vcpu} cores, respectively, and produce a 37\% error for memory, and 
    -5\% and -6\% of error, respectively, for \ac{lc}.
    Noteworthy, all models achieve an error close to 0\% for \ac{lc}, posing a significant finding considering the substantial impact of \ac{lc} on the performance of the \ac{vfw}. However, negative errors by the benchmarks indicate under-provisioning predictions compared to the required \ac{lc}. The over-provisioning predictions made by the \ac{rl} model are preferred over the under-provisioning exhibited by the benchmarks, as the latter results in sub-optimal \ac{or} performance of the \ac{vfw}.    

\begin{figure*}[h]
    \centering 
    \begin{subfigure}{0.475\columnwidth}
        \includegraphics[width=\textwidth]{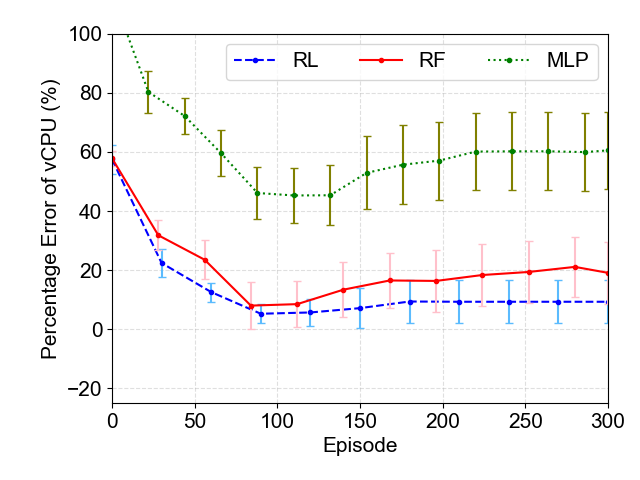}
        \caption{}
        \label{fig:error_firewall_a}
    \end{subfigure}
    \begin{subfigure}{0.45\columnwidth}
        \includegraphics[width=\textwidth]{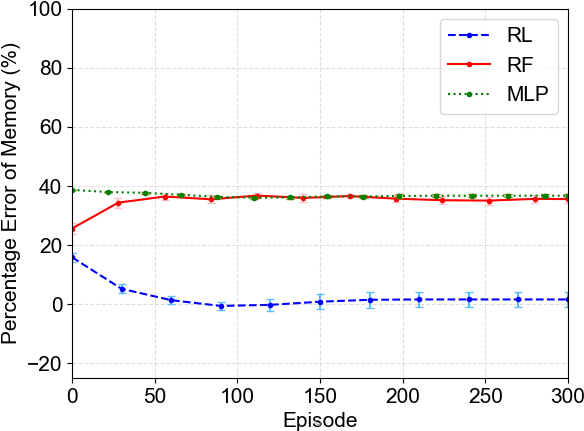}
        \caption{}
        \label{fig:error_firewall_b}
    \end{subfigure}
    \begin{subfigure}{0.45\columnwidth}
        \includegraphics[width=\textwidth]{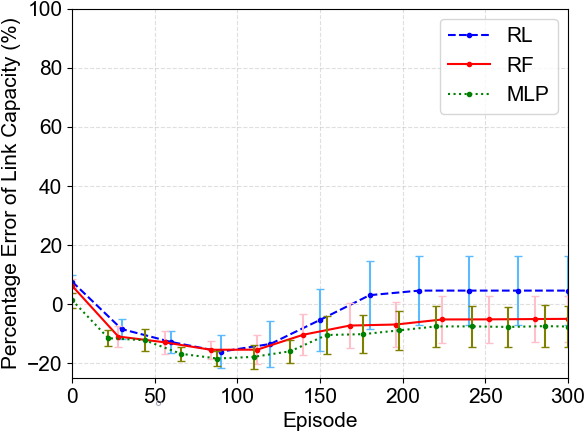}
        \caption{}
        \label{fig:error_firewall_c}
    \end{subfigure}
    \caption{Percentage error of resource allocation predictions by \ac{mlp}, \ac{rf} and \ac{rl} regarding the \vfw \ac{vnf}.} 
    \label{fig:predict_vFW}
\end{figure*}



\subsection{Resource optimisation scenarios} 
\label{sec:combine_weight}

We examine the impact of resource objectives on performance as a result of the resource type importance on the optimisation problem.
The latter 
is captured via weighted parameters in the scalarised Q-Learning equation of formula~(\ref{eq:sq_linear}). 
We investigate \textit{39 scenarios} (13 per \ac{vnf} type) with different \ac{rl} resource allocation objective weights, with our findings presented in Tables~\ref{table:output_steady_state_inline}, \ref{table:output_steady_state_passive}, and \ref{table:output_steady_state_vfw} for each type of \ac{vnf}.

\subsubsection{Snort (Inline mode)}

Our findings in Tab.\,\ref{table:output_steady_state_inline} show that all weight configurations can reduce \ac{vcpu} core usage to around 40\%. Specifically, setting the weight of \ac{vcpu} to one ($w(1,0,0)$) leads to a 40\% reduction in \ac{vcpu} usage, but also results in an 80\% reduction in memory usage and a low link utilisation of 34.38\% (as expressed by $\sfrac{OR}{LC}$). On the other hand, weight configurations such as $w(\sfrac{1}{2},\sfrac{1}{2},0)$ and $w(\sfrac{1}{2},0,\sfrac{1}{2})$ increase \ac{vcpu} usage, reduce memory usage, and increase link utilisation to almost 80\% respectively.
Our analysis reveals that the weight of \ac{lc} has a higher impact on \ac{vcpu} usage than the weight of memory reward. Furthermore, the weight of \ac{vcpu} and the weight of \ac{lc} do not affect memory usage. The steady-state LC utilisation for $w(0,0,1)$, $w(0,\sfrac{1}{2},\sfrac{1}{2})$, and $w(\sfrac{1}{2},0,\sfrac{1}{2})$ is 92.73, 91.70, and 79.54 respectively. If the weight of \ac{vcpu} is increased, the OR/LC decreases significantly, while the weight of memory has no significant effect on the OR/LC.
Finally, compared to schemes with high resource weighting or equal weighting, $w(\sfrac{1}{2},\sfrac{1}{2},0)$, which only weights \ac{vcpu} and memory, does not increase link utilisation. As a result, considering the weight of \ac{lc} is critical in enhancing LC utilisation.
%
\begin{table}[h]
\centering
\caption{Mean steady-state output for \snort (Inline mode) in 13 resource-scalarisation scenarios.}
\label{table:output_steady_state_inline}
 \begin{tabular}{|l|c|c|c|c|c|c|c|}
    \hline
    \rowcolor[HTML]{C0C0C0}  \multicolumn{2}{|l|}{\cellcolor[HTML]{C0C0C0}} & vCPU & MEM & LC & \begin{tabular}[c]{@{}c@{}}vCPU\\ (2 cores)\end{tabular} & \begin{tabular}[c]{@{}c@{}}MEM\\ (1.6 GB)\end{tabular} & OR/LC (\%) 
    \\ \hline
    & vCPU & 1   &     &     & 40\% & 100.00\%  & 34.38 \\ \cline{2-8} 
    & MEM &     & 1   &     & 90\% & 68.81\%  & 91.62 \\ \cline{2-8} 
    \multirow{-3}{*}{Single Resource Weighted} 
    & LC &     &     & 1   & 90\% & 100.00\%  & 92.73 
    \\ \hline
    & vCPU \& MEM & $\sfrac{1}{2}$ & $\sfrac{1}{2}$ &     & 44\% & 69.94\% & 41.25 \\ \cline{2-8} 
    & MEM \& LC  &     & $\sfrac{1}{2}$ & $\sfrac{1}{2}$ & 90\% & 69.69\% & 91.70 \\ \cline{2-8} 
    \multirow{-3}{*}{Two Resource Weighted} 
    & vCPU \& LC & $\sfrac{1}{2}$ &     & $\sfrac{1}{2}$ & 50\% & 100.00\% & 79.54 
    \\ \hline
    \multicolumn{1}{|c|}{} 
    & & $\sfrac{3}{4}$ & $\sfrac{1}{8}$ & $\sfrac{1}{8}$ & 41\% & 68.00\% & 59.82 \\ \cline{3-8} 
    \multicolumn{1}{|c|}{} & \multirow{-2}{*}{vCPU} 
    & $\sfrac{1}{2}$ & $\sfrac{1}{4}$ & $\sfrac{1}{4}$ & 41\% & 67.75\% & 57.87 \\ \cline{2-8} 
    \multicolumn{1}{|c|}{}                                         
    & & $\sfrac{1}{8}$ & $\sfrac{3}{4}$ & $\sfrac{1}{8}$ & 42\% & 68.38\% & 58.97 \\ \cline{3-8} 
    \multicolumn{1}{|c|}{}& \multirow{-2}{*}{MEM}                       
    & $\sfrac{1}{4}$ & $\sfrac{1}{2}$ & $\sfrac{1}{4}$ & 41\% & 68.25\% & 59.95 \\ \cline{2-8} 
    \multicolumn{1}{|c|}{}                                         
    & & $\sfrac{1}{8}$ & $\sfrac{1}{8}$ & $\sfrac{3}{4}$ & 42\% & 68.00\% & 60.78 \\ \cline{3-8} 
    \multicolumn{1}{|c|}{\multirow{-6}{*}{High Resource Weighted}} & 
    \multirow{-2}{*}{LC} & $\sfrac{1}{4}$ & $\sfrac{1}{4}$ & $\sfrac{1}{2}$ & 41\% & 67.62\% & 60.00 
    \\ \hline
    Equal Weights & vCPU \& MEM \& LC & $\sfrac{1}{3}$ & $\sfrac{1}{3}$ & $\sfrac{1}{3}$ & 41\% & 66.31\% & 52.20 
    \\ \hline
    \end{tabular}
\end{table}

\subsubsection{Snort (Passive mode)}

    The case of only \ac{vcpu} $w(1, 0, 0)$ in Tab.\,\ref{table:output_steady_state_passive} minimises demand for \ac{vcpu} cores to 40\%. \ac{vcpu} usage can be reduced to 44\% and 42\% also in the case of $w(\sfrac{1}{2}, \sfrac{1}{2}, 0)$ and $w(\sfrac{1}{2}, 0, \sfrac{1}{2})$, respectively. 
         Memory weight has a  slight impact on allocated \ac{vcpu} cores.
         Memory and \ac{lc} weights have a minor impact on allocating \ac{vcpu} cores. 
         The weight of \ac{vcpu} has more impact on memory allocation than that of \ac{lc}. 
         The weight of \ac{vcpu} has a greater influence on the \ac{lc} utilisation than the memory weight.
         In conclusion, 
         \snort Passive mode with $w(\sfrac{1}{3}, \sfrac{1}{3}, \sfrac{1}{3})$ can effectively reduce resource usage while achieving a high link utilisation OR/LC. 
%

\subsubsection{Virtual FireWall}

        The need for \ac{vcpu} cores is reduced to 40\% for the case of $w(1, 0, 0)$. 
    In the cases of $w(\sfrac{1}{2}, \sfrac{1}{2}, 0)$ and $w(\sfrac{1}{2}, 0, \sfrac{1}{2})$, \ac{vcpu} usage can be decreased to 42\% and 40\%, respectively. 
    In this case, \ac{lc} and memory weights have no impact on the allocation of \ac{vcpu} cores. 
        Memory allocation for the cases of $w(0, 1, 0)$, $w(\sfrac{1}{2}, \sfrac{1}{2}, 0)$, and $w(0, \sfrac{1}{2}, \sfrac{1}{2})$ is reduced to 55\%, 56\%, and 56\%, respectively. 
In this case, the weight of \ac{vcpu} and \ac{lc} does not impact memory.
        For weight combinations $w(0, 0, 1)$, $w(0, \sfrac{1}{2}, \sfrac{1}{2})$, and $w(\sfrac{1}{2}, 0, \sfrac{1}{2})$, link utilisation OR/LC in steady state is 94.45\%, 94.64\%, and 88.84\%, respectively.
    Consequently, link utilisation is influenced more by the weight of \ac{vcpu} than by the weight of memory. 
        Regarding all high resource-weighted cases, steady-state \ac{vcpu} and memory were about 39\% and 55\%, respectively. However, for $w(\sfrac{1}{2}, \sfrac{1}{4}, \sfrac{1}{4})$ we observe the highest link utilisation  \ac{or}/\ac{lc}, almost 90\%, which is about 9\% higher than the case of the high weight of the \ac{lc} (around 81\%).
         To conclude, \ac{vfw} with $w(\sfrac{1}{2}, \sfrac{1}{4}, \sfrac{1}{4})$ can cut down on resource consumption while offering high OR/LC.
\begin{table}[]
\centering
\caption{Mean steady-state output for \snort (Passive mode) in 13 resource-scalarisation scenarios.}
\label{table:output_steady_state_passive}
    \begin{tabular}{|l|c|c|c|c|c|c|c|}
    \hline
    \rowcolor[HTML]{C0C0C0}  \multicolumn{2}{|l|}{\cellcolor[HTML]{C0C0C0}} & vCPU & MEM & LC  & \begin{tabular}[c]{@{}c@{}}vCPU\\ (2 cores)\end{tabular} & \begin{tabular}[c]{@{}c@{}}MEM\\ (1.6 GB)\end{tabular} & OR/LC (\%) 
    \\ \hline
    & vCPU & 1   &     &     & 40\% & 100.00\%  & 61.62 \\ \cline{2-8} 
    & MEM &     & 1   &     & 90\% & 68.81\%  & 81.38 \\ \cline{2-8} 
    \multirow{-3}{*}{Single Resource Weighted} 
    & LC &     &     & 1   & 90\% & 100.00\% & 92.46 
    \\ \hline
    & vCPU \& MEM & $\sfrac{1}{2}$ & $\sfrac{1}{2}$ &     & 44\% & 72.06\% & 61.88 \\ \cline{2-8} 
    & MEM \& LC  &     & $\sfrac{1}{2}$ & $\sfrac{1}{2}$ & 90\% & 70.44\% & 88.34 \\ \cline{2-8} 
    \multirow{-3}{*}{Two Resource Weighted} 
    & vCPU \& LC & $\sfrac{1}{2}$ &     & $\sfrac{1}{2}$ & 42\% & 100.00\% & 68.65 
    \\ \hline
    \multicolumn{1}{|c|}{} 
    & & $\sfrac{3}{4}$ & $\sfrac{1}{8}$ & $\sfrac{1}{8}$ & 34\% & 66.25\% & 58.19 \\ \cline{3-8} 
    \multicolumn{1}{|c|}{} & \multirow{-2}{*}{vCPU}
    & $\sfrac{1}{2}$ & $\sfrac{1}{4}$ & $\sfrac{1}{4}$ & 33\% & 65.00\% & 54.01 \\ \cline{2-8} 
    \multicolumn{1}{|c|}{}                                         
    & & $\sfrac{1}{8}$ & $\sfrac{3}{4}$ & $\sfrac{1}{8}$ & 35\% & 66.44\% & 60.00 \\ \cline{3-8} 
    \multicolumn{1}{|c|}{} & \multirow{-2}{*}{MEM}
    & $\sfrac{1}{4}$ & $\sfrac{1}{2}$ & $\sfrac{1}{4}$ & 33\% & 65.19\% & 54.41 \\ \cline{2-8} 
    \multicolumn{1}{|c|}{}                                         
    & & $\sfrac{1}{8}$ & $\sfrac{1}{8}$ & $\sfrac{3}{4}$ & 34\% & 67.00\% & 57.61 \\ \cline{3-8} 
    \multicolumn{1}{|c|}{\multirow{-6}{*}{High Resource Weighted}} & \multirow{-2}{*}{LC}
    & $\sfrac{1}{4}$ & $\sfrac{1}{4}$ & $\sfrac{1}{2}$ & 34\% & 65.50\% & 56.72 
    \\ \hline
    Equal Weight & vCPU \& MEM \& LC 
    & $\sfrac{1}{3}$ & $\sfrac{1}{3}$ & $\sfrac{1}{3}$ & 35\% & 67.12\% & 60.64 
    \\ \hline
    \end{tabular}
\end{table}

%
\begin{table}[h]
\centering
\caption{Mean steady-state output for \vfw in 13 resource-scalarisation scenarios.}
\label{table:output_steady_state_vfw}
    \begin{tabular}{|l|c|c|c|c|c|c|c|c|}
    \hline
    \rowcolor[HTML]{C0C0C0} \multicolumn{2}{|l|}{\cellcolor[HTML]{C0C0C0}}  & vCPU & MEM & LC  & \begin{tabular}[c]{@{}c@{}}vCPU\\ (2 cores)\end{tabular} & \begin{tabular}[c]{@{}c@{}}MEM\\ (1.6 GB)\end{tabular} & OR/LC (\%) 
    \\ \hline
    & vCPU & 1   &     &     & 40\% & 100.00\%  & 92.25 \\ \cline{2-8} 
    & MEM &     & 1   &     & 90\% & 68.81\%  & 92.50 \\ \cline{2-8} 
    \multirow{-3}{*}{Single Resource Weighted} 
    & LC &     &     & 1   & 90\% & 100.00\% & 94.45 
    \\ \hline
    & vCPU \& MEM & $\sfrac{1}{2}$ & $\sfrac{1}{2}$ &     & 42\% & 70.06\% & 87.50 \\ \cline{2-8} 
    & MEM \& LC  &     & $\sfrac{1}{2}$ & $\sfrac{1}{2}$ & 90\% & 69.69\% & 94.64 \\ \cline{2-8} 
    \multirow{-3}{*}{Two Resource Weighted} 
    & vCPU \& LC & $\sfrac{1}{2}$ &     & $\sfrac{1}{2}$ & 40\% & 100.00\% & 88.84 
    \\ \hline
    \multicolumn{1}{|c|}{} 
    & & $\sfrac{3}{4}$ & $\sfrac{1}{8}$ & $\sfrac{1}{8}$ & 39\% & 68.75\% & 85.20 \\ \cline{3-8} 
    \multicolumn{1}{|c|}{}& \multirow{-2}{*}{vCPU}
    & $\sfrac{1}{2}$ & $\sfrac{1}{4}$ & $\sfrac{1}{4}$ & 40\% & 68.50\% & 89.10 \\ \cline{2-8} 
    \multicolumn{1}{|c|}{}                                         
    & & $\sfrac{1}{8}$ & $\sfrac{3}{4}$ & $\sfrac{1}{8}$ & 39\% & 68.62\% & 83.44 \\ \cline{3-8} 
    \multicolumn{1}{|c|}{} & \multirow{-2}{*}{MEM} 
    & $\sfrac{1}{4}$ & $\sfrac{1}{2}$ & $\sfrac{1}{4}$ & 39\% & 68.31\% & 84.91 \\ \cline{2-8} 
    \multicolumn{1}{|c|}{}                                         
    & & $\sfrac{1}{8}$ & $\sfrac{1}{8}$ & $\sfrac{3}{4}$ & 40\% & 69.81\% & 80.75 \\ \cline{3-8} 
    \multicolumn{1}{|c|}{\multirow{-6}{*}{High Resource Weighted}} & \multirow{-2}{*}{LC} 
    & $\sfrac{1}{4}$ & $\sfrac{1}{4}$ & $\sfrac{1}{2}$ & 39\% & 68.38\% & 82.45 
    \\ \hline
    Equal Weight & vCPU \& MEM \& LC 
    & $\sfrac{1}{3}$ & $\sfrac{1}{3}$ & $\sfrac{1}{3}$ & 39\% & 69.75\% & 79.57 
    \\ \hline
    \end{tabular}
\end{table}

\subsubsection{ {\Revised Highlight conclusions \& limitations}}

Link Capacity is more untactful on \ac{or} performance in \snort Passive mode than in \snort Inline mode and \ac{vfw} because traffic is forwarded directly to the destination without being inspected before forwarding. 
\ac{vfw} gives the highest \ac{or} to \ac{lc} ratio at around 83.63\% compared to \snort Inline mode (58.50\%) and \snort Passive mode (57.36\%). 
Because \ac{vfw} drops packets incoming to unallowed ports and forwards packets from allowed ports, packet delay does not occur in this \ac{vnf}. But unlike \ac{vfw}, \snort Passive duplicates packets with a latency stop before forwarding them, and \snort Inline packets must be inspected before being sent to the output link. This inspection delay causes \textit{congestion in the output link}.
When considering the effect of the weights of each resource's reward function on reducing the corresponding resource
while maintaining the \ac{or}, we find that the 
behaviour of each \ac{vnf} is different. 

{\Revised 
Finally, we acknowledge the following experimental limitations.
First, the performance of \ac{vfw} and Snort can fluctuate under a constant resource allocation, depending upon the number of configuration rules loaded into the system. Our method assumes that all configurable aspects of VNF behaviour, aside from resource allocation, exhibit relative stability throughout the  \ac{vnf}'s lifecycle. Future research should assess the performance of \ac{rl} in scenarios involving dynamic configuration changes.
Second, iPerf has limited traffic generation capabilities, e.g., it struggles to reach rates $\geq$1~Gbps, and packets are not completely realistic. While a well-configured iPerf suffices for demonstrating the proposed method and showcasing an experimental proof of concept, 
it cannot fully capture a production network deployment with real traffic.}

\section{Conclusion} 
\label{sec:conclusion}

{\Revised
We introduce iOn-Profiler as an intelligent \textit{online learning} \ac{vnf} profiler using \ac{ml} and in particular \ac{rl}, incorporating Q-Learning across a range of optimisation objectives. This autonomous profiling \ac{rl} model-based adjusts to network dynamics and our work and study results demonstrate its effectiveness by improving the efficiency of profiling for two modes of the Snort (Inline mode and Passive mode) \ac{vnf} and for \ac{vfw}, a virtual firewall \ac{vnf}.
%
%
We investigate 39 scenarios (13 per \ac{vnf} type) with different \ac{rl} resource allocation objective weights to understand the impact of different resource types on the quality of our profiling model's resource allocation decisions. 
Our comprehensive evaluation results 
highlight the importance of considering multiple resource optimisation objectives and examining each \ac{vnf} type individually with online learning, rather than with a statically trained \ac{sl} model that is impossible to adapt to dynamics such as in demand patterns or be easily used for transfer learning purposes.

Our future research plans include expanding our model for \textit{service function chains} in various \ac{vnf} configurations and for different \ac{vnf} types across different network resource substrates. We also aim to further explore network adaptability and the benefits of iOn-Profiler with \textit{transfer learning} between different resources and/or across different \ac{vnf} types.
}

\section*{Acknowledgement}
 This work received funding from UK funded Project REASON under the FONRC sponsored DSIT, and EU projects 5G-VICTORI and 5GASP (grant agreements No. 857201 and No. 101016448).

\bibliographystyle{ieeetr}
{ \bibliography{references} }

\begin{thebibliography}{10}

\bibitem{vios_pro}
R.~Nejabati, S.~Moazzeni, P.~Jaisudthi, and D.~Simenidou, ``Zero-touch network orchestration at the edge,'' in {\em 30th International Conference on Computer Communications and Networks, {ICCCN} 2021, Athens, Greece, July 19-22, 2021}, pp.~1--5, {IEEE}, 2021.

\bibitem{NAP}
S.~Moazzeni, P.~Jaisudthi, A.~Bravalheri, N.~Uniyal, X.~Vasilakos, R.~Nejabati, and D.~Simeonidou, ``A novel autonomous profiling method for the next-generation nfv orchestrators,'' {\em IEEE Transactions on Network and Service Management}, vol.~18, no.~1, pp.~642--655, 2020.

\bibitem{snort_modes}
T.~Morris, R.~Vaughn, and Y.~Dandass, ``A retrofit network intrusion detection system for modbus rtu and ascii industrial control systems,'' in {\em 2012 45th Hawaii International Conference on System Sciences}, pp.~2338--2345, IEEE, 2012.

\bibitem{luizelli2015piecing}
M.~C. Luizelli, L.~R. Bays, L.~S. Buriol, M.~P. Barcellos, and L.~P. Gaspary, ``Piecing together the nfv provisioning puzzle: Efficient placement and chaining of virtual network functions,'' in {\em 2015 IFIP/IEEE International Symposium on Integrated Network Management (IM)}, pp.~98--106, IEEE, 2015.

\bibitem{fang2016}
W.~Fang, M.~Zeng, X.~Liu, W.~Lu, and Z.~Zhu, ``Joint spectrum and it resource allocation for efficient vnf service chaining in inter-datacenter elastic optical networks,'' {\em IEEE Communications Letters}, vol.~20, no.~8, pp.~1539--1542, 2016.

\bibitem{ztorch}
V.~Sciancalepore {\em et~al.}, ``z-torch: An automated nfv orchestration and monitoring solution,'' {\em IEEE Transactions on Network and Service Management}, vol.~15, no.~4, pp.~1292--1306, 2018.

\bibitem{montgomery2021introduction}
D.~C. Montgomery, E.~A. Peck, and G.~G. Vining, {\em Introduction to linear regression analysis}.
\newblock John Wiley \& Sons, 2021.

\bibitem{gou2019generalized}
J.~Gou, H.~Ma, W.~Ou, S.~Zeng, Y.~Rao, and H.~Yang, ``A generalized mean distance-based k-nearest neighbor classifier,'' {\em Expert Systems with Applications}, vol.~115, pp.~356--372, 2019.

\bibitem{verma2019interpolation}
V.~Verma, K.~Kawaguchi, A.~Lamb, J.~Kannala, Y.~Bengio, and D.~Lopez-Paz, ``Interpolation consistency training for semi-supervised learning,'' {\em arXiv preprint arXiv:1903.03825}, 2019.

\bibitem{xu2021applying}
A.~Xu, H.~Chang, Y.~Xu, R.~Li, X.~Li, and Y.~Zhao, ``Applying artificial neural networks (anns) to solve solid waste-related issues: A critical review,'' {\em Waste Management}, vol.~124, pp.~385--402, 2021.

\bibitem{juliano2020nonlinear}
S.~A. Juliano, ``Nonlinear curve fitting: predation and functional response curves,'' in {\em Design and analysis of ecological experiments}, pp.~159--182, Chapman and Hall/CRC, 2020.

\bibitem{profile_RA}
S.~Van~Rossem {\em et~al.}, ``Profile-based resource allocation for virtualized network functions,'' {\em IEEE Transactions on Network and Service Management}, vol.~16, no.~4, pp.~1374--1388, 2019.

\bibitem{bunyakitanon2020auto}
M.~Bunyakitanon, A.~P. {da Silva}, X.~Vasilakos, R.~Nejabati, and D.~Simeonidou, ``{Auto-3P: An autonomous VNF performance prediction \& placement framework based on machine learning},'' {\em Computer Networks}, vol.~181, p.~107433, 2020.

\bibitem{2020:are3p}
M.~{Bunyakitanon}, X.~{Vasilakos}, R.~{Nejabati}, and D.~{Simeonidou}, ``{End-to-End Performance-based Autonomous VNF Placement with adopted Reinforcement Learning},'' {\em {IEEE Transactions on Cognitive Communications and Networking}}, pp.~1--1, 2020.

\bibitem{iprofiler}
P.~Jaisudthi, S.~Moazzeni, X.~Vasilakos, R.~Nejabati, and D.~Simeonidou, ``i-profiler: Towards multi-objective autonomous vnf profiling with reinforcement learning,'' in {\em IEEE INFOCOM 2023 - IEEE Conference on Computer Communications Workshops (INFOCOM WKSHPS)}, pp.~1--6, 2023.

\bibitem{iglesias2017orca}
J.~O. Iglesias, J.~A. Aroca, V.~Hilt, and D.~Lugones, ``Orca: an orchestration automata for configuring vnfs,'' in {\em Proceedings of the 18th ACM/IFIP/USENIX middleware conference}, pp.~81--94, 2017.

\bibitem{mestres2018machine}
A.~Mestres {\em et~al.}, ``A machine learning-based approach for virtual network function modeling,'' in {\em 2018 IEEE Wireless Communications and Networking Conference Workshops (WCNCW)}, pp.~237--242, IEEE, 2018.

\bibitem{Manuel2018}
M.~Peuster and H.~Karl, ``Understand your chains and keep your deadlines: Introducing time-constrained profiling for nfv,'' in {\em 2018 14th International Conference on Network and Service Management (CNSM)}, pp.~240--246, IEEE, 2018.

\bibitem{van2020vnf}
S.~Van~Rossem, W.~Tavernier, D.~Colle, M.~Pickavet, and P.~Demeester, ``{VNF performance modelling: From stand-alone to chained topologies},'' {\em Computer Networks}, vol.~181, p.~107428, 2020.

\bibitem{khan2018nfv}
M.~G. Khan, S.~Bastani, J.~Taheri, A.~Kassler, and S.~Deng, ``{NFV-Inspector: A systematic approach to profile and analyze virtual network functions},'' in {\em 2018 IEEE 7th international conference on cloud networking (CloudNet)}, pp.~1--7, IEEE, 2018.

\bibitem{ravin}
V.~R. Chintapalli, V.~S.~K. Giduturi, B.~R. Tamma, and A.~Antony~Franklin, ``{RAVIN: A Resource-aware VNF Placement Scheme with Performance Guarantees},'' in {\em NOMS 2023-2023 IEEE/IFIP Network Operations and Management Symposium}, pp.~1--9, 2023.

\bibitem{vnf-prof}
N.~Ferdosian, S.~Moazzeni, P.~Jaisudthi, Y.~Ren, H.~Agrawal, D.~Simeonidou, and R.~Nejabati, ``{Autonomous Intelligent VNF Profiling for Future Intelligent Network Orchestration},'' {\em IEEE Transactions on Machine Learning in Communications and Networking}, pp.~1--1, 2023.

\bibitem{perfo-bottleneck}
R.~Jia, H.~Pan, H.~Jiang, S.~Fdida, and G.~Xie, ``Towards diagnosing accurately the performance bottleneck of software-based network function implementation,'' in {\em Passive and Active Measurement}, (Cham), pp.~227--253, Springer Nature Switzerland, 2023.

\bibitem{profile_chains}
M.~Peuster and H.~Karl, ``Profile your chains, not functions: Automated network service profiling in devops environments,'' in {\em 2017 IEEE Conference on Network Function Virtualization and Software Defined Networks (NFV-SDN)}, pp.~1--6, IEEE, 2017.

\bibitem{yilma2020benchmarking}
G.~M. Yilma, Z.~F. Yousaf, V.~Sciancalepore, and X.~Costa-Perez, ``{Benchmarking open source NFV MANO systems: OSM and ONAP},'' {\em Computer communications}, vol.~161, pp.~86--98, 2020.

\bibitem{helicon}
M.~Bunyakitanon, X.~Vasilakos, R.~Nejabati, and D.~Simeonidou, ``{HELICON:} orchestrating low-latent {\&} load-balanced virtual network functions,'' in {\em {IEEE} International Conference on Communications, {ICC} 2022, Seoul, Korea, May 16-20, 2022}, pp.~353--358, {IEEE}, 2022.

\bibitem{uniyal21}
N.~Uniyal, A.~Bravalheri, X.~Vasilakos, R.~Nejabati, D.~Simeonidou, W.~Featherstone, S.~Wu, and D.~Warren, ``Intelligent mobile handover prediction for zero downtime edge application mobility,'' in {\em {IEEE} Global Communications Conference, {GLOBECOM} 2021, Madrid, Spain, December 7-11, 2021}, pp.~1--6, {IEEE}, 2021.

\end{thebibliography}

\end{document}